\documentclass[12pt,a4paper]{article}
\usepackage{graphicx}
\usepackage{a4wide}
\usepackage{amsmath}

\begin{document}
\thispagestyle{empty}
 
\title{Platonic topology  and CMB fluctuations:\\ 
Homotopy, anisotropy,  and multipole selection rules.}
\author{Peter Kramer,\\ Institut f\"ur Theoretische Physik der Universit\"at  T\"ubingen,\\ Germany.}
\maketitle

%%corrections 12/2/2010, 16/4/2010
\section*{Abstract.}

The Cosmic Microwave Background CMB originates from an early stage in the history of the universe. Observations show  low multipole contributions of CMB fluctuations.
A possible explanation is given by a non-trivial topology of the universe and has motivated 
the search for topological selection rules.
Harmonic analysis on a topological manifold must provide  basis sets for all functions compatible with a given topology and so is needed to model the CMB fluctuations. 
We analyze the fundamental groups of  Platonic tetrahedral, cubic, and octahedral manifolds 
using deck transformations.
From them we construct the appropriate harmonic analysis and boundary conditions. We provide the algebraic means for modelling the multipole expansion of incoming CMB radiation. From the assumption of randomness we derive  selection rules, depending on the point symmetry of the manifold.

\section{Introduction.}

Temperature fluctuations of the incoming Cosmic Microwave Background, originating  from an early state of the universe, are being studied with great precision. We refer to \cite{HI09} for the data. The weakness of the observed amplitudes for low multipole orders, see for example \cite{COP05}, was taken by various authors  as a motivation to explore constraints on the multipole amplitudes coming from non-trivial topologies of the universe. The topology of the universe is not fixed by the differential equations of Einstein's theory of gravitation.
It has been suggested that specific topological models lead to the selection and exclusion of certain low multipole moments and might explain the observations. Harmonic analysis provides a basis set for all functions compatible with a given topology. The general idea is to model the amplitude of incoming CMB radiation by these basis functions. Since the observations provide multipole expansion in two angular coordinates, we discuss the reduction from the three variables on the 3-sphere to the observer frame.

The authors of \cite{LEV02}, \cite{LA95} survey general topological concepts and give particular  examples applied to cosmology.
For representative  contributions by other authors, we quote \cite{LEV02}, 
\cite{LE02}, \cite{LU03}, \cite{AU05}, \cite{AU05b}, \cite{AU08}, \cite{WE04}, \cite{BE06}, \cite{WE07}, \cite{CAI07}, and 
references given in these papers. In contrast to the previous numerical work, we give analytical expressions for the basis sets. As most of these authors we assume average positive curvature
and so analyze spherical 3-manifolds.

Spherical 3-manifolds come in families, one of particular interest being the class of Platonic polyhedra (see i.e.\ the dodecahedron in \cite{WE04,KR05}). Here we present a systematic approach to the Platonic polyhedra, based on new insight into the homotopy \cite{EV04}, which enables us to characterize new topologies. In the present work, we construct for these topologies the harmonic analysis and provide the algebraic tools for modelling the multipole expansion of CMB fluctuations.

In section~\ref{sec:fromhomotopy}  we discuss our general methods. In section  \ref{sec:thethreesphere} we set the coordinates and introduce the Wigner polynomials.
In section \ref{platonicthreemanifolds} we describe   the harmonic analysis on seven spherical Platonic 3-manifolds. We demonstrate and illuminate in section~\ref{sec:thetwospher}
the map from  homotopies to deck transformations for cubic 3-manifolds and the homotopic boundary conditions.
In section~\ref{sec:modellingincoming} we set the algebraic tools for the modelization of the CMB and then prove multipole selection rules for randomly chosen functions.
The novel points of the present analysis are summarized in section~\ref{sec:summary}.
Three Appendices deal with \ref{sec:synopsisof} the details of seven Platonic manifolds, \ref{sec:Wigner} the Wigner polynomials, and 
\ref{sec:fromdeckvia}  random point symmetry.

\section{From homotopy  to harmonic analysis on a manifold.}
\label{sec:fromhomotopy}

We now examine how the fundamental group determines the analysis
of functions on a manifold.
The topology of a manifold ${\cal M}$ is characterized by its homotopy. For general notions of topology 
we refer to \cite{SE34} and \cite{TH97}.
The  fundamental or first homotopy group $\pi_1({\cal M})$ of a manifold ${\cal M}$ has as its elements inequivalent classes of continuous paths on the manifold ${\cal M}$,  returning to the same point.
Group multiplication is given by path concatenation. 

The simply connected universal cover of a manifold offers another view. This cover, for spherical 3-manifolds the 3-sphere $S^3$, is tiled
by copies of ${\cal M}$. The tiling is produced by the fixpoint-free action of a group $H$.
The discrete groups acting fixpoint-free on covers are called space forms. The general classification of spherical space forms of low dimension is given by Wolf \cite{WO84} 
pp.~224-226.

For cosmic topology, these  groups acting on $S^3$ are taken as the starting point in \cite{LE02}
and \cite{AU05}. These and other authors choose actions of these groups on
$S^3$ and compute by numerical methods  \cite{AU05}, or with the help of the Laplacian \cite{LE02}, corresponding eigenmodes. It is assumed that the groups $H$ are isomorphic to
the fundamental group of a manifold. 

We do not follow this route since there is no unique pathway from the bare group $H$ back  to 
the manifold ${\cal M}$, its geometric boundaries and  homotopies. 
First of all, from eq.~\ref{6}, and as noted in \cite{LE02} pp.~4686-4687, the unitary unimodular group $SU(2,C)$, and so any of its discrete subgroups, admits
at least three different types of actions on $S^3$.
Moreover, 
cyclic groups $H=C_n$, denoted in \cite{LE02} as $Z_n$ and associated there on pp.~4688 with lens space manifolds, emerge   in our analysis as $C_5$ for the tetrahedral manifold $N1$ and 
as $C_8$ for the cubic manifold $N2$. It follows that a bare cyclic group  acting on $S^3$ does  not determine a unique topological manifold.

Instead we follow \cite{SE34} and take the manifold and its fundamental group of homotopies as our starting point.
The group $H$ that generates the tiling on the cover we call with \cite{SE34} the group
of deck transformations $H={\rm deck}({\cal M})$.
Homotopy and deck transformations are linked together in a theorem due to Seifert and Threlfall 
\cite{SE34} pp.~196-198. 
It  states that, for a manifold ${\cal M}$, the group of deck transformations and 
the fundamental group of homotopies are isomorphic.

For polyhedral manifolds, homotopies are generated by the gluing of boundaries. 
Everitt in \cite{EV04} determines (with minor corrections given  in \cite{CAV09}), all possible homotopies of the family of Platonic 
polyhedra on $S^3$.
The theorem by Seifert and Threlfall ensures  that from any  face gluing generator of a polyhedral  homotopy one can find  an associated  deck transformation
that maps on $S^3$ a prototile to a neighbouring image tile. 
We implement this theorem and construct, on the basis  of  work detailed in \cite{KR05}, \cite{KR08}, \cite{KR09}, \cite{KR09b}, from homotopic face gluings 
the isomorphic generators of deck transformations.
In this way we derive from homotopies, rather than postulate, the individual groups $H$ of deck transformations and at the same time obtain  definite  actions of them on $S^3$.

The constructed group $H={\rm deck}({\cal M})$ is our key to the analysis of functions on the manifold.
Harmonic analysis on $S^3$, with the domain being a spherical  3-manifold, is the exact spherical counterpart 
of Euclidean crystallographic Fourier series analysis. In fact, the analysis of Euclidean 
cosmic topology given in \cite{AU08} exemplifies   crystallographic  analysis for topology 
with Euclidean cover. 
On $S^3$, we start from the Wigner polynomials since they span an orthonormal  harmonic basis for square integrable functions. By algebraic projection and multiplicity analysis we construct for each Platonic manifold all  $H$-invariant linear combinations of Wigner polynomials. These in turn  span a basis that respects the tiling, incorporates 
the fundamental group, and has the  spherical manifold as its domain. From our analysis there follow strict boundary conditions, set by homotopy, on pairs of faces
of the spherical polyhedron.

\section{The 3-sphere, its isometries, and Wigner polynomial bases.}
\label{sec:thethreesphere}

Our starting point for the functional analysis on spherical manifolds are the 3-sphere,
its coordinates,  
and an orthogonal system of harmonic polynomials  on it.
The points of the 3-sphere are in one-to-one correspondence to the elements of the unitary unimodular group $SU(2,C)$. This allows to choose coordinates on the 3-sphere. We label them by a $2 \times 2$ unimodular matrix in the form
\begin{equation}
\label{1} 
u=\left[
\begin{array}{ll}
 z_1&z_2\\
-\overline{z}_2&z_1\\
\end{array}
\right],\: z_1=x_0-ix_3,\: z_2=-x_2-ix_1,\: z_1\overline{z}_1+z_2\overline{z}_2=1.
\end{equation}
where the real coordinates of $E^4$ are $x=(x_0,x_1,x_2,x_3)$.
The group of isometries of $S^3$ is $SO(4,R)$. This group can be expressed as the direct product of two groups $SU^l(2,C), SU^r(2,C)$ in the form
\begin{equation}
\label{2}
SO(4,R) \sim  (SU^l(2,C)\times  SU^r(2,C))/Z_2.
\end{equation}
The action of these groups on $u\in S^3$ is given by
\begin{equation}
\label{3}
(g_l,g_r)\in  (SU^l(2,C)\times  SU^r(2,C)): u \rightarrow g_l^{-1}ug_r.
\end{equation}
The subgroup $Z_2$ in eq.~\ref{2} is generated by $(g_l, g_r)=(-e,-e)\in (SU^l(2,C)\times  SU^r(2,C))$.

The diagonal subgroup in  eq.~\ref{2} with elements $(g,g)\in  (SU^l(2,C)\times  SU^r(2,C))$ we denote by $SU^C(2,C)$. 
The actions of $SU^C(2,C)$ on $u$ produce rotations $R(g)$ wrt.\ the three coordinates $(x_1,x_2,x_3)$.
Any element of $SO(4,R)$ can be uniquely factorized  as 
\begin{eqnarray}
\label{4} 
&&(g_l,g_r')= (g_l,g_l) (e,g_r),\: g_r:=g_l^{-1}g_r',
\\ \nonumber  
&&(g_l,g_l)\in SU^C(2,C),\:  (e,g_r) \in SU^r(2,C).
\end{eqnarray}
These relations  express the fact that  the points of the coset space $SO(4,R)/SU^C(2,C)$ can be identified with  the elements of $SU^r(2,C)$, and hence with the points of the 3-sphere $S^3$. 
The 4-dimensional spherical harmonics have this coset space as their domain. As shown in \cite{KR08},
these spherical harmonics can be identified with the Wigner $D^j$-functions
\begin{equation}
 \label{4a}
D^j_{m_1,m_2}(u), \: 2j=0,1, \ldots, \infty,\: -j\leq (m_1,m_2)\leq j.
\end{equation}
Wigner \cite{WI59} pp.~166-170 introduced them  as the unitary irreducible representations of the group $SU(2,C)$, 
often parametrized by three Euler angles eq.~\ref{a1}.
Since the Wigner $D^j$ functions can be seen as a complete orthogonal system of polynomial functions on $S^3$, homogeneous of degree $2j$ in the four complex matrix elements of $u$, see Appendix \ref{sec:Wigner}, we coin for them the name Wigner polynomials. As shown in \cite{KR08} Lemma 5 p. 3526, these polynomials are harmonic, i.e. vanish under application of the Laplacian in
$E^4$, eq.~\ref{A3}.\\ 
The action of a general element  $(g_l,g_r)\in  (SU^l(2,C)\times  SU^r(2,C))$ on a spherical harmonic eq.~\ref{4a} 
is given, using the representation properties of $D^j$,  by
\begin{eqnarray}
\label{5}
&& (T_{(g_l,g_r)}D^j_{m_1,m_2})(u) := D^j_{m_1,m_2}(g_l^{-1}ug_r)
\\ \nonumber 
&&= \sum_{(m_1', m_2')=-j}^j D^j_{m_1'm_2'}(u)\: \left[D^j_{m_1,m_1'}(g_l^{-1})D^j_{m_2',m_2}(g_r)\right],
\end{eqnarray}
in terms of products of two Wigner $D^j$-functions of the acting group elements $(g_l, g_r)$.
It follows that the 3-sphere supports  both the action of the group $H$ and the $H$-invariant basis of the harmonic analysis.
Moreover, it allows to compare spherical topologies with different homotopies, groups $H$, and harmonic analysis.

\subsection{Projection and multiplicity for $H$-invariant polynomials.} 
Given the group $H$ of deck transformations, we can project on the chosen manifold 
a basis for the harmonic analysis. The projector $P^0$ from a Wigner to a $H$-invariant polynomial produces, using eq.~\ref{5}, the linear combination
\begin{equation}
\label{5b}
(P^0D^j_{m_1m_2})(u)=  \sum_{m_1'm_2'} D^j_{m_1'm_2'}(u)\left[\frac{1}{|H|}\sum_{(g_l,g_r)\in H}\: D^j_{m_1m_1'}(g_l^{-1})D^j_{m_2'm_2}(g_r)\right]. 
\end{equation}
By standard methods of group representations, the multiplicity $m(j,0)$ of linear independent $H$-invariant polynomials for given $j$ is, from  computing the characters of the
representation eq.~\ref{5}, given by 
\begin{equation}
 \label{5c}
m(j,0)= \frac{1}{|H|}\sum_{(g_l,g_r)\in H} \chi^j(g_l^{-1})\chi^j(g_r),
\end{equation}
with $\chi^j(g)$ the character \cite{WI59} pp.~155-156 of $g \in SU(2,C)$ and $|H|$ the order of
$H$. 
The multiplicity eq.~\ref{5c} controls the number of linearly independent projections
eq.~\ref{5b}.

\subsection{Boundary conditions for the harmonic analysis set by the homotopy group $H$.}
\label{sec:boundaryconditions}

For a given polyhedral shape, the first homotopy group $H$ is generated by the gluing of pairs of faces.
The isomorphic map of homotopies to deck transformations is sketched in section~\ref{sec:fromhomotopy} and carried out in our previous work. Now pairs of faces glued by homotopy appear in the tiling generated by deck transformations as boundaries shared by pairs of neighbouring polyhedral tiles. This map and the boundary conditions are demonstrated in  \ref{sec:thetwospher}.
There  follows

\paragraph{Prop 1:}  Any $H$-invariant polynomial, defined on the polyhedron, must repeat its values on pairs of faces of the prototile linked by the elements of $H= {\rm deck}({\cal M})$ isomorphic to the  homotopic gluing of faces.

The harmonic analysis on the polyhedral prototile therefore is subject to these boundary conditions.
Homotopies from the same polyhedral shape are distinguished by their boundary conditions.
Moreover, since the underlying Wigner polynomials are harmonic, we have

\paragraph{Prop 2:} The $H$-invariant polynomials on a polyhedron solve the Laplace equation inside 
the polyhedron. Their values are repeated on pairs of faces, related by  the face gluings from the 
group $H$ of homotopies.

\section{Platonic 3-manifolds, groups of deck transformations, and bases for the harmonic analysis.}\label{platonicthreemanifolds}

\subsection{Coxeter groups on the 3-sphere and Platonic polyhedra.}

\begin{table}
\begin{tabular}{|l|l|l|l|l|l|}\hline
Coxeter diagram $\Gamma$ & $|\Gamma|$ & Polyhedron ${\cal M}$ & $H={\rm deck}({\cal M})$ & $|H|$ & Reference \\ \hline
$\circ -\circ -\circ - \circ$               & $120$  & tetrahedron $N1$ & $C_5$         & $5$ & \cite{KR08} \\  \hline
$\circ \stackrel{4}{-} \circ -\circ -\circ$ & $384$  & cube $N2$        & $C_8$         & $8$ & \cite{KR09} \\                                                           &        & cube $N3$        & $Q$           & $8$ &\cite{KR09} \\  \hline
$\circ -\circ \stackrel{4}{-}\circ - \circ$ & $1152$ & octahedron $N4$  & $C_3\times Q$ & $24$ &\cite{KR09b} \\
                                            &        & octahedron $N5$  & $B$           & $24$ &\cite{KR09b} \\
                                            &        & octahedron $N6$  & ${\cal T}^*$  & $24$&\cite{KR09b} \\ \hline
$\circ -\circ -\circ \stackrel{5}{-} \circ$ & $120\cdot 120$ & dodecahedron $N1'$ & ${\cal J}^*$  & $120$ & \cite{KR05}, \cite{KR06} \\ \hline
\end{tabular}
\caption{\label{Table1}
4 Coxeter groups $\Gamma$, 4 Platonic polyhedra ${\cal M}$, 7 groups $H={\rm deck}({\cal M})$ of order $|H|$.
In the Table, $C_n$ denotes a cyclic, $Q$ the quaternion, ${\cal T}^*$ the binary tetrahedral, ${\cal J}^*$ the binary icosahedral group. The symbols $Ni$ are adapted from  \cite{EV04}.}
\end{table}

To construct the Platonic 3-manifolds we follow \cite{EV04} and introduce Coxeter groups $\Gamma$ generated by 
reflections in hyperplanes of $E^4$. One reason for their use is that all faces  of
the Platonic polyhedra are located on such reflection hyperplanes. Moreover the Platonic tilings 
of $S^3$ can be found from   the defining representations on $E^4$ of these groups.
We shall use in subsection \ref{sec:representations} the representations of the Coxeter groups to construct the deck transformations.
In section \ref{sec:modellingincoming} we shall use these Coxeter groups to discuss the random point symmetry of the manifolds.
 
Given Euclidean space with standard scalar product $\langle, \rangle$, a Weyl reflection $W_a$ with unit Weyl vector 
$a: \langle a, a\rangle=1$ acts 
on $x\in E^4$ as 
\begin{equation}
 \label{6} 
W_a: x \rightarrow  W_a\: x:=x-2\langle x, a \rangle\: a,\: (W_a)^2=e.
\end{equation}
This   is a reflection in the hyperplane perpendicular to the unit vector $a$.
Coxeter groups $\Gamma$ \cite{HU90} are generated by Weyl reflections with relations of the type $(W_{a_i} W_{a_j})^{m_{ij}}=e$.
The Coxeter diagram encodes Weyl reflections  by nodes, and by integer numbers $m_{ij}$ for relations of pairs of generating Weyl reflections. A horizontal line in the diagram between two nodes denotes the particular  value $m_{ij}=3$.
The numbers $m_{ij}$ determine also the scalar products between the corresponding pairs of Weyl unit vectors by 
\begin{equation}
 \label{6b}
\langle a_i, a_j\rangle =\cos(\frac{\pi}{m_{ij}}).
\end{equation}
Pairs of unlinked nodes in the Coxeter diagram yield Weyl reflection vectors perpendicular 
to one another and reflections that commute.

The Platonic polyhedra in 3 dimensions form a family of regular polyhedra,  bounded by the regular 2-dimensional polygons. Similarly, 
the m-cells, see  \cite{SO58}, are a family of regular polyhedra in 4 dimensions, bounded by the regular 3-dimensional Platonic polyhedra.
By projection of the Euclidean geometric objects to the spheres $S^2$ and $S^3$ respectively, one obtains
spherical polygons, polyhedra and m-cells. The geometric symmetry of these objects we express in terms 
of 4 Coxeter reflection groups $\Gamma$ . Their diagrams are given in Table~\ref{Table1} and their  four generators in Table~\ref{Table2}. For a fixed Coxeter group we use the short-hand notation $\Gamma: W_{a_j}=:W_i$.

The Weyl reflection planes of the first three generators of $\Gamma$ pass through the point 
$(1,0,0,0)$ and bound a cone.
The intersection of this cone with the  Weyl reflection plane of the fourth generator bounds what is called the Coxeter simplex. 
Each of the Coxeter groups $\Gamma$ tiles $S^3$ into $|\Gamma|$ copies of a fundamental Coxeter simplex. 
In topology we are interested in actions preserving orientation. The maximal subgroup of a Coxeter group with this 
property is generated by the products $(W_1W_2),(W_2W_3),(W_3W_4)$ of generators. Its representation on $E^4$ is given by 
unimodular matrices with determinant $1$.  Because of its unimodular representation we denote this subgroup
by $S\Gamma$, and find for its order $|S\Gamma|=|\Gamma|/2$. The fundamental domain for $S\Gamma$ can be taken as 
a duplex, formed by a mirror pair of Coxeter simplices.

The Platonic polyhedra are built from sets of Coxeter simplices sharing a single vertex, as illustrated in Figs. 2-8. 

The group $H= {\rm deck}({\cal M})$ is a subgroup of $\Gamma$ and produces on $S^3$ a second, superimposed tiling by $|H|$ copies of a Platonic polyhedron ${\cal M}$. Since $|H|$ must be equal to $m$, the tiling is a $|H|$-cell on $S^3$. The Platonic $|H|$-cells are discussed and illustrated  in  \cite{SO58}.

\begin{table}
\begin{equation*}
\begin{array}{|l|l|l|l|l|} \hline
\Gamma & a_1 & a_2 &a_3& a_4\\ \hline
\circ -\circ -\circ - \circ &  (0,0,0,1)&(0,0,\sqrt{\frac{3}{4}},\frac{1}{2})
& (0,\sqrt{\frac{2}{3}},\sqrt{\frac{1}{3}},0)& (\sqrt{\frac{5}{8}},\sqrt{\frac{3}{8}},0,0)\\ \hline
\circ \stackrel{4}{-} \circ -\circ -\circ&  (0,0,0,1)& (0,0,-\sqrt{\frac{1}{2}},\sqrt{\frac{1}{2}})
&(0,\sqrt{\frac{1}{2}},-\sqrt{\frac{1}{2}},0)&(-\sqrt{\frac{1}{2}},\sqrt{\frac{1}{2}},0,0)\\ \hline
\circ -\circ \stackrel{4}{-}\circ - \circ& (0,\sqrt{\frac{1}{2}},-\sqrt{\frac{1}{2}},0)&(0,0,-\sqrt{\frac{1}{2}},\sqrt{\frac{1}{2}})
&(0,0,0,1)& (\frac{1}{2},\frac{1}{2},\frac{1}{2},\frac{1}{2})\\ \hline
\circ -\circ -\circ \stackrel{5}{-} \circ& (0,0,1,0)&(0,-\frac{\sqrt{-\tau+3}}{2},\frac{\tau}{2},0)
&(0,-\sqrt{\frac{\tau+2}{5}},0,-\sqrt{\frac{-\tau+3}{5}})&(\frac{\sqrt{2-\tau}}{2},0,0,-\frac{\sqrt{\tau+2}}{2}) \\ \hline
\end{array}
\end{equation*}
\caption{\label{Table2}
The Weyl vectors $a_s$  
for the four Coxeter groups $\Gamma$ from Table~\ref{Table1} with  $\tau:=\frac{1+\sqrt{5}}{2}$.
}
\end{table}

The Platonic polyhedra become topological 3-manifolds upon specifying fundamental groups or homotopies for them as is done in \cite{EV04}. 
We adopt the notation $Nj$ for these manifolds. Note from section \ref{sec:thetwospher} that a single polyhedral  shape can carry several 
inequivalent fundamental groups of equal order. For a list of the  non-abelian groups of order $\leq 30$
we refer to \cite{CO65} pp.~134-135, Table~1. Binary symmetry groups, given as subgroups of $SU(2,C)$, we denote by a star $^*$.

\subsection{Representation of products of Weyl reflections.}
\label{sec:representations}

We shall construct the deck transformations of the Platonic polyhedra from 
even products of Weyl reflections. Here we provide the appropriate algebraic tools. 
For a Weyl reflection with Weyl unit vector $a_s$, we define the $2 \times 2$ matrix $v_s$ by
inserting the four Cartesian components of $a_s$ into eq.~\ref{1}. The product $(W_{a_i}W_{a_j})$ is a rotation 
in $E^4$. The corresponding rotation operator from \cite{KR08} eq.(60)  can be written in terms of  the matrices $(v_i, v_j)$ as
\begin{equation}
\label{8}
T_{(W_{a_i}W_{a_j})}= T_{(v_iv_j^{-1},v_i^{-1}v_j)},
\end{equation}

and for fixed degree $2j$ has the representation given in eq.~\ref{5}.
All deck transformations appearing in what follows are orientation-preserving and therefore must
be products of an even number of Weyl reflections. Eq.~\ref{8} guarantees that all of them can be expressed 
by pairs $(g_l,g_r)$.

\section{The two spherical cubic manifolds.}
\label{sec:thetwospher}

\begin{figure}
\begin{center}
\includegraphics[width=1.0\textwidth]{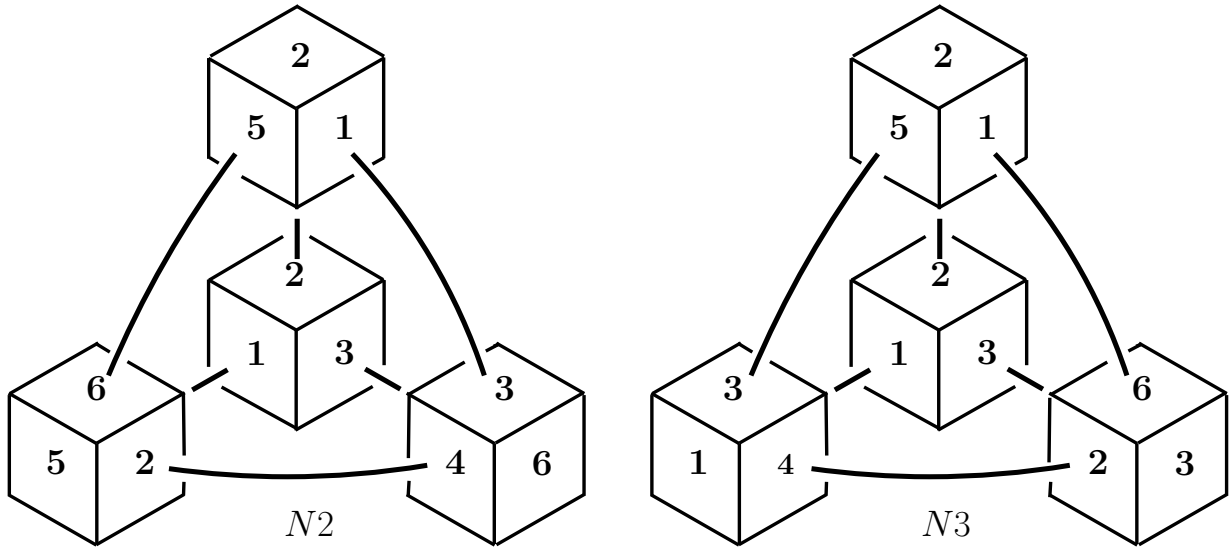} 
\end{center}
\caption{\label{fig:ncubes} The cubic manifolds $N2$ and $N3$. The  cubic prototile and three neighbour tiles sharing 
its faces $F1,F2,F3$. The four cubes are replaced by their  Euclidean counterparts  and separated from one another. Visible faces are denoted by the numbers from Fig.~\ref{fig:cubus2}. The actions transforming the prototile into its three neighbours generate the deck transformations and the 8-cell tiling of $S^3$. In the tiling, homotopic face gluing takes the form of  shared
pairs of faces 
$N2: F3\cup F1, F4\cup F2, F6 \cup F5$ and $N3: F1\cup F6, F2\cup F4, F3 \cup F5$. It is  marked by heavy lines or arcs.}
\end{figure}

In this section we illustrate the use of the cover $S^3$ and of deck transformations by two spherical
cubic 3-manifolds. They are the spherical counterparts of the Euclidean cubic manifold
discussed in \cite{AU08}. Everitt in \cite{EV04} constructs two inequivalent cubic manifolds which we denote by $N2,N3$. His face and edge gluings are given in section \ref{subsec:thecubic}. These we illustrate in Fig.~\ref{fig:ncubes}. All spherical cubes are replaced by their Euclidean 
counterparts. We start from the cubic prototile and the homotopic gluing of its faces, enumerated as $F1, F2,F3$ according to Fig.~\ref{fig:cubus2}. These gluings we transform in \ref{subsec:thecubic} into deck transformations from the prototile to its three neighbour tiles, shown separated with parallel faces in the Figure. When passing to $S^3$, the Euclidean cubes are replaced by spherical cubes
of the cubic 8-cell tiling of $S^3$, see \cite{SO58} p. 177-8 and \cite{KR09} Fig.~1.
The enumeration of the visible faces in the figure shows that they have been rotated. 
In the 8-cell tiling of $S^3$, the homotopic gluing appears as the sharing of faces. In the figure we connect pairs of  shared faces  by heavy 
lines or arcs. The differences between the left $N2$ and right $N3$ manifold illustrates the two inequivalent fundamental groups found in \cite{EV04}, both  with the same cubic shape of the prototile. 
The corresponding groups $H$ of deck transformations we construct algebraically in  \ref{subsec:thecubic} as a cyclic group
$H=C_8$ for $N2$ and the quaternion group $H=Q$ for $N3$, both of order $8$. 
The bases of the harmonic analysis on the two cubic manifolds we find by algebraic projection 
as $H$-invariant linear combinations of Wigner polynomials. They are 
listed in subsection \ref{subsec:thecubic}. Their values on the prototile differ in their homotopic boundary conditions.
Therefore we expect for them in general different anisotropies and multipole expansions.

\section{Modelling incoming CMB by harmonic analysis.}
\label{sec:modellingincoming}

In this section we discuss  the algebraic tools for analysing incoming  CMB radiation 
in terms of the harmonic bases for a chosen topology.

\subsection{Alternative coordinates on $S^3$.}
For the harmonic analysis on spherical 3-manifolds we use the spherical harmonics in the form of 
Wigner polynomials. These polynomials in the coordinates $x$ are often expressed in terms of 
Euler angle coordinates, Edmonds \cite{ED57} pp.~53-67:
\begin{equation}
\label{a1}
\begin{array}{|l|} \hline
 x_0= \cos(\frac{\alpha+\gamma}{2})\cos(\frac{\beta}{2}),
\: x_1= -\sin(\frac{\alpha-\gamma}{2})\sin(\frac{\beta}{2}),
\\ 
 x_2=  -\cos(\frac{\alpha-\gamma}{2})\sin(\frac{\beta}{2}),
\: x_3= -\sin(\frac{\alpha+\gamma}{2})\cos(\frac{\beta}{2}),
\\ \hline
\\
 u=
\left[
\begin{array}{ll}
 \exp(\frac{i(\alpha+\gamma)}{2})\cos(\frac{\beta}{2}),& \exp(\frac{i(\alpha-\gamma)}{2})\sin(\frac{\beta}{2})\\
 -\exp(\frac{i(-\alpha+\gamma)}{2})\sin(\frac{\beta}{2}),&  \exp(\frac{-i(\alpha+\gamma)}{2})\cos(\frac{\beta}{2})
\end{array}
\right]\\ 
\\ \hline
\end{array}
\end{equation}
We give the coordinates and the form of the matrix $u$ eq.~\ref{1}.
An alternative system of polar coordinates is used by Aurich et al.~\cite{AU05}. Here 
\begin{equation}
\label{a2}
\begin{array}{|l|} \hline
 x_0= \cos(\chi), \:x_1= \sin(\chi)\sin(\theta)\cos(\phi),
\\ 
 x_2=\sin(\chi)\sin(\theta)\sin(\phi),\: x_3= \sin(\chi)\cos(\theta),
\\ \hline
\\
 u=
\left[
\begin{array}{ll}
 \cos(\chi)-i\sin(\chi)\cos(\theta),& -i\sin(\chi)\sin(\theta)\exp(-i\phi)\\
 -i\sin(\chi)\sin(\theta)\exp(i\phi),    & \cos(\chi)+i\sin(\chi)\cos(\theta)
\end{array}
\right]\\ 
\\ \hline
\end{array}
\end{equation}
We shall see in eq.~\ref{a7} that these polar coordinates are adapted to the analysis of incoming radiation in terms of 
its direction.
\subsection{Multipole expansion of  spherical harmonics on $S^3$.}
For a clear description of the multipole analysis of the CMB we refer to \cite{AU05}.
We relate our analysis algebraically to this description.
The Wigner polynomials eq.~\ref{A1} in Euler angle coordinates eq.~\ref{a1}
from \cite{ED57} p.~55 factorize as
\begin{equation}
\label{a3}
 D^j_{m_1,m_2}(u)= \exp(im_1\alpha)d^j_{m_1,m_2}(\beta)\exp(im_2\gamma)
\end{equation}
To adapt  the Wigner polynomials to a multipole expansion, we transform them 
for fixed degree $2j$ by use of Wigner coefficients of $SU(2,C)$, \cite{ED57} pp.~31-45, into the new harmonic polynomials
\begin{eqnarray}
\label{a4}
 && \psi_{\beta l m}(u) = \delta_{\beta,2j+1}
\sum_{m_1,m_2}D^j_{m_1,m_2}(u) \langle j-m_1jm_2|lm\rangle (-1)^{j-m_1},
\\ \nonumber
&& l=0,1,..., 2j=\beta-1.
\end{eqnarray}
This transformation links the Wigner polynomials to the basis given  in \cite{AU05} whose index notation we adopt.
Whereas the index $j$ of the Wigner polynomials can be integer or half-integer, the multipole index $l$ takes only 
integer values. For fixed $l$ we have $2j\geq l$, and for fixed $2j$:  $0\leq l\leq 2j$.
Using representation theory of $SU(2)$ it can be shown from eqs. \ref{5} and \ref{a4} that the conjugation action $u \rightarrow g^{-1}ug$ of the group $SU^C(2,C)$ acts by a rotation $R(g)$  only on the coordinate triple $(x_1,x_2,x_3)$, and  the new 
polynomials eq.~\ref{a4} transform as 
\begin{equation}
\label{a5}
(T_{(g,g)}\psi_{\beta l m})(u)=\psi_{\beta l m}(g^{-1}ug) = \sum_{m'=-l}^l\psi_{\beta l m'}(u)D^l_{m',m}(g),
\end{equation}
like the spherical harmonics $Y^l_m(\theta,\phi)$. We therefore adopt eq.~\ref{a5} as the action of the usual rotation group for 
cosmological models covered by the 3-sphere. Eq.~\ref{a5} qualifies $l$ as the multipole index of incoming radiation.

The basis transformation eq.~\ref{a4} can be inverted with the help of the orthogonality of the Wigner coefficients \cite{ED57} to yield
\begin{equation}
\label{a6}
 D^j_{m_1,m_2}(u)= \delta_{\beta, 2j+1} \delta_{m,-m_1+m_2}\sum_{l=0}^{2j}\psi_{\beta l m}(u) \langle j-m_1jm_2|lm\rangle (-1)^{j-m_1}
\end{equation}
The result eq.~\ref{a5} can be further elaborated  by  use of the alternative coordinates $(\chi, \theta, \phi)$  eq.~\ref{a2}.
We follow Aurich et al. \cite{AU05}, eqs. 9-17, to find
 \begin{eqnarray}
\label{a7}
 &&\psi_{\beta l m}(u)= R_{\beta l}(\chi)Y^l_m(\theta,\phi),
\\ \nonumber
&&R_{\beta l}(\chi)=2^{l+\frac{1}{2}}l!\sqrt{\frac{\beta(\beta-l-1)!}{\pi(\beta+l)}}
C^{l+1}_{\beta-l-1}(\cos(\chi))
\end{eqnarray}
where $C^{l+1}_{\beta-l-1}$ is a Gegenbauer polynomial. A similar expression is given in 
\cite{LE02} pp. 4705-7.
Eq.~\ref{a7} shows that the alternative spherical harmonics eq.~\ref{a4}, written in the polar coordinates 
eq.~\ref{a2}, admit the separation into a part depending on $\chi$ and a standard spherical harmonic as a function of polar 
coordinates $(\theta, \phi)$. 
For a  very clear interpretation  of the role of the coordinate $\chi$, appearing in the Gegembauer polynomials of eq.~\ref{a7}, its relation to cosmological models, and to the surface of last scattering, we refer to  \cite{AU05}.

\subsection{Harmonic analysis and anisotropy from spherical manifolds.}

Observed  anisotropies of the CMB fluctuations are discussed for example in  \cite{ZH08} and \cite{SC09}.
From the present point of view,
there are two different sources of anisotropy in the harmonic analysis, which apply to Platonic as well as to other 
polyhedral topologies. 

\subsection{Anisotropy  from the orientation of the polyhedron.}

Although the 3-sphere is isotropic with respect to rotations, any  polyhedral  prototile 
has a particular orientation, chosen with the Weyl reflection vectors with respect to the frame of coordinates $x=(x_0,x_1,x_2,x_3)$.
 For any model derived from a spherical  topological manifold, it follows that frames  of
different orientation on $S^3$  must be explored independent from one another.
There is no motivation for  averaging. The most general rotation of the frame of coordinates transforms 
the Wigner polynomials of fixed degree $2j$ according to eq.~\ref{5}. 

\subsection{Anisotropy from the underlying homotopy group.}

One way to model the CMB by a given Platonic 3-manifold is to  combine its $H$-invariant basis  polynomials, ordered by degree $2j$, linearly with random coefficients, pass with the transformation eq.~\ref{a6} and coordinate transformation \ref{a2} to the new basis eq.~\ref{a7}, 
consider the dependence  on $\chi$ mentioned after eq.~\ref{a7}, 
and evaluate the resulting multipole expansion.

We argue that this general procedure does not ensure simple  selection rules for the multipole expansion. 
The reason is that the full basis  must strictly obey the boundary conditions on pairs of faces found from homotopy in section~\ref{sec:boundaryconditions}.

Evidence   for the impact  of homotopies on the basis of the harmonic analysis is provided in Appendix~\ref{sec:synopsisof}  by the tetrahedral manifold $N1$, the cubic manifolds $N2, N3$, and by the octahedral manifold $N4$: In all these cases we find new preferred  
coordinate settings  $x' \sim u'$ such that 
the $H$-invariant   bases become  very simple linear combinations of Wigner polynomials $D^j(u')$. These particular coordinate settings from  homotopy must produce observable anisotropies.

\subsection{Random polyhedral point symmetry and multipole selection rules.}
\label{sec:randompolyhedral}

Homotopy implies  boundary conditions in the harmonic analysis for pairs of polyhedral faces. These conditions  are much weaker than those implied by the geometric rotational point group $M\in SO(3,R)$ of symmetries of the polyhedron.
Conversely, the boundary conditions from homotopy do not exclude the geometrical point symmetry  of the polyhedron. 
The compatibility of the point and the deck groups is discussed in Appendix~\ref{sec:fromdeckvia}. We now show that under an additional assumption of randomness there follow multipole selection rules 
of the type which motivated the search for non-trivial topologies.

$M$-invariance restricts the domain of a function on a polyhedron to a conal domain of a volume fraction 
$\frac{1}{|M|}$. 
For any $M$-invariant function we have:

\paragraph{Prop 3:} If a function, defined on a regular polyhedron, is invariant under its point symmetry group $M$,
it also fulfills the boundary conditions from any of its homotopy groups.

{\em Proof}: The action of $M$ on the faces of a regular polyhedron is transitive, i.e. transforms   any pair of faces
into one another. It also contains the polyhedral rotations preserving the midpoint
of any face. Therefore  it follows from $M$-invariance that the  boundary values of the function on the faces do obey  any homotopic boundary conditions as discussed in section~\ref{sec:boundaryconditions}.

Among possible functions with domain the polyhedron,
consider now  a random function $\Psi^{{\rm random}}(u)$. From eq.~\ref{a4}, any point group element $R(h) \in M$ acts on this random function as 
\begin{equation}
\label{aa8}
(x_1,x_2,x_3) \rightarrow R(h)(x_1,x_2,x_3),\: u \rightarrow h^{-1}uh,\: h\in SU^C(2,C)
.\end{equation}
This rotation by assumption preserves the geometrical shape of the manifold, and on it produces a new admissible random function
\begin{equation}
\label{aa9}
\Psi^{{\rm random}}(u)\rightarrow (T_{(h,h)} \Psi^{{\rm random}})(u)=\Psi^{{\rm random}}(h^{-1}uh)
\end{equation}
The values of a proper  random function  on a polyhedron with geometrical symmetry group $M$  should not distinguish between different 
orientations eqs. \ref{aa8} and \ref{aa9} within the same geometrical  shape.  
It follows that the two random functions eq.~\ref{aa8} and eq.~\ref{aa9} must coincide. Applying this argument to all elements of $M$ it follows that  the random function $\Psi^{{\rm random}}(u)$  must be $M$-invariant with domain the conical section 
described before Prop 3.

Now we can infer selection rules for the multipole
expansion of this random function. For use in molecular physics, the relation between   point 
symmetry and total rotational angular momentum is well  studied. Listed  
for example in \cite{LA74}, pp.~436-438, is the multiplicity $m(l,\downarrow \Gamma_p)$  of the representation $\Gamma_p, p=1,2,...$ of the point group $M$ of a molecule contained in the representation $D^l, l=0,1,2,...$ of the rotation group. From  Frobenius reciprocity, see \cite{COL68} p. 86,
it follows that the multiplicity $m(\Gamma_p\uparrow  l)$ of linearly independent functions, constructed 
from a function belonging to the representation $\Gamma_p$ of the point group $M$ and  transforming under rotations according to  $D^l$, obeys
\begin{equation}
\label{m5}
 m(\Gamma_p \uparrow l)=m(l \downarrow\Gamma_p).
\end{equation}
This rule applies in particular to the identity representation $\Gamma_1$ of the point group $M$.
The random function $\Psi^{{\rm random}}(u)$ eq.~\ref{aa9} is assumed to be $M$-invariant  and so belongs to the
representation $\Gamma_1$ of $M$. Application of eq.~\ref{m5} gives

\paragraph{Prop 4:} A random function $\Psi^{\rm random}(u)$ on a (spherical) polyhedral topological 3-manifold,
invariant under its  point group $M$, can contribute to the multipole order $l$ only if $m(l\downarrow  \Gamma_1) \geq 1$.

A direct proof of this proposition follows by use of eqs. \ref{a5}, \ref{a7}: For given multipole order $l$, 
the projector to the identity representation $\Gamma_1$ of the rotational polyhedral symmetry group $M$
acts only on the spherical harmonics $Y^l_m(\theta,\phi)$. It gives a non-vanishing result only if
$m(l\downarrow  \Gamma_1) \geq 1$.
In the Table~\ref{Table61}, adapted from \cite{LA74}, we collect the relevant numbers $m(l\downarrow \Gamma_1)$ for some point groups
up to multipole order $l=6$. Recursive results for higher values of $l$ are given in the same reference.

\begin{table}
\begin{equation*}
 \begin{array}{|l||l|l|l|l|l|l|l|l|l|l|l|} \hline
l& {\cal C}_2&D_2&{\cal C}_3&D_3&{\cal C}_4&D_4&{\cal C}_6&D_6&T&O&{\cal J}\\ \hline
0&1&1&1&1&1&1&1&1&1&1&1\\ \hline
1&1&0&1&0&1&0&1&0&0&0&0\\ \hline
2&3&2&1&1&1&1&1&1&0&0&0\\ \hline 
3&3&1&3&1&1&0&1&0&1&0&0\\ \hline
4&5&3&3&2&3&2&1&1&1&1&0\\ \hline
5&5&2&3&1&3&1&1&0&0&0&0\\ \hline
6&7&4&5&3&3&2&3&2&2&1&1\\ \hline
\end{array}
\end{equation*}
\caption{\label{Table61}
The multiplicity $m(\Gamma_1\downarrow l)$ for the values $l,\: 0\leq l\leq 6$ of the multipole order and selected point groups $M$ in the notation of \cite{LA74}, assuming a function invariant under $M$. Most numbers are from  \cite{LA74} pp.~436-438, the last column from eq.~\ref{j1}.
}
\end{table}

Clearly the assumption of random polyhedral point symmetry, combined with homotopy, yields strong multipole selection rules. 
For the Platonic polyhedral 3-manifolds studied here we find:
The tetrahedron has  lowest multipole orders $l=0,3,4,6^2$, the cube and octahedron lowest multipole orders $l=0, 4,6$,
the dodecahedron and icosahedron lowest multipole orders $l=0,6$. In Appendix~\ref{sec:fromdeckvia} we exemplify the 
deck and point groups and the onset of invariant polynomials 
for the cubic manifold $N3$, and in section \ref{sec:thetetrahedral}  we give selection rules 
from point symmetry for the tetrahedral manifold $N1$.

\section{Summary.}
\label{sec:summary}

We summarize here  the salient points of the present work, which to our knowledge are not covered in  the work on cosmic topology published by other authors:

(1) {\bf Platonic topologies}: We deal mainly  with the family of Platonic spherical 3-manifolds whose homotopies have recently been  derived in \cite{EV04}. 
Harmonic analysis on these  manifolds with homotopies given in \cite{EV04} is not available from other sources. A great deal
of our general  methods 
apply to non-Platonic  polyhedral 3-manifolds.

(2) {\bf Start from the fundamental group}: The starting point taken for each spherical 3-manifold is its fundamental or first homotopy group. 
We remove any ambiguity in the group action by always starting from the geometry and the fundamental group of the polyhedral 
manifold. The only  remaining freedom is the orientation of the 
quadruple of Weyl vector for the associated Coxeter group $\Gamma$. This freedom must be explored as the frame dependence
of the modelization, point (7).
By an  elementwise rigorous conversion of  homotopy groups  we construct the isomorphic group $H$ of deck transformations. On this basis we  derive  left, right, or two-sided actions of $H$ on $S^3$. Our distinction of these actions agrees with the one used in \cite{LE02}.

(3) {\bf Inequivalent topologies from a single polyhedron}:
The work \cite{EV04} lists inequivalent homotopy groups for a chosen Platonic polyhedron.
We follow  \cite{EV04} and  give for spherical cubes   two inequivalent groups, illustrated in \ref{sec:thetwospher},   for spherical 
octahedra three inequivalent groups $H$ of homotopies, isomorphic groups of deck transformations, and bases for the harmonic analysis. The harmonic bases differ in their homotopic boundary conditions.

(4) {\bf Algebraic harmonic analysis and homotopic boundary conditions}: 
The harmonic analysis is developed on the universal cover $S^3$. We use the Wigner  harmonic polynomials, Appendix~\ref{sec:Wigner}, which form 
a complete orthonormal basis on the domain $S^3$.
The basis for the harmonic analysis on a spherical  manifold is constructed 
by most other authors in the field by numerical methods, see for example \cite{AU05} p.9 or \cite{LU03}. For the Platonic 3-manifolds  we always proceed algebraically  by use of  group representations. The bases  are spanned by the  $H$-invariant subsets of Wigner polynomials on the 3-sphere. In particular for the manifolds $N1-N4$, Wigner polynomials give extremely   simple results.

We show in section~\ref{sec:boundaryconditions} that the basis functions of the harmonic analysis obey  boundary conditions on pairs of polyhedral faces and so reflect the chosen homotopy.

(5) {\bf Group/subgroup analysis}: The selection rules for a specific 3-manifold we illuminate by  representations of groups intermediate between 
the rotation group $O(4,R)$ and the specific group $H$ of deck transformations. We put the group $H$,  
the Coxeter group $\Gamma$, and its unimodular subgroup $S\Gamma$  into the subgroup relation $H<S\Gamma<SO(4,R)$.
Selection rules from the representations of these groups  we derive 
in particular 
for the tetrahedral manifold $N1$, see \cite{KR08} and Table~\ref{TableN1b}, and for the two cubic manifolds $N2, N3$, see \cite{KR09}. Even stronger selection rules result from
the assumption of random point symmetry in Appendix \ref{sec:fromdeckvia}.

(6) {\bf Algebraic multipole analysis}: By an  algebraic transformation,  combined with a transformation of angular coordinates given in  sections 6.1-2,
we adapt  the Wigner polynomials to  an explicit multipole expansion with standard transformation properties eq.~\ref{a5} 
under rotations, as used in observing the CMB.

(7) {\bf Anisotropy}: 
We point out two sources of anisotropy. The first one comes from  the orientation of the polyhedral prototile,
the second one, exemplified by the tetrahedral, cubic and octahedral manifolds, reflects the boundary conditions of the harmonic analysis set by homotopy.

(8) {\bf From random point symmetry to multipole selection rules}:
We show in section~\ref{sec:randompolyhedral} that the additional assumption of  random geometrical polyhedral point symmetry, in conjunction with homotopy of the polyhedral manifold, implies strong multipole selection rules for CMB radiation. We emphasize   that similar selection rules from deck and
random point symmetry  apply to regular polyhedral topologies of hyperbolic and Euclidean type.

\section*{Acknowledgment.}
The author appreciates substantial  general advice  and help in algebraic computations by Dr. Tobias Kramer,
Institut f\"ur Theoretische Physik der Universit\"at Regensburg, Germany.

\appendix

\section{Synopsis of Platonic polyhedral manifolds.}
\label{sec:synopsisof}

In this section  we illustrate in figures the polyhedra in relation to Coxeter groups and the enumeration of faces and edges, elaborate  for the seven spherical Platonic spherical 3-manifolds listed in Table~\ref{Table1},  
the homotopy in terms of  face and edge gluings, the groups $H$  of deck transformations and their action on $S^3$, and the basis 
for the harmonic analysis in terms of Wigner polynomials.

\subsection{The tetrahedral manifold $N1$.}
\label{sec:thetetrahedral}
\begin{figure}
\begin{center}
\includegraphics[width=0.5\textwidth]{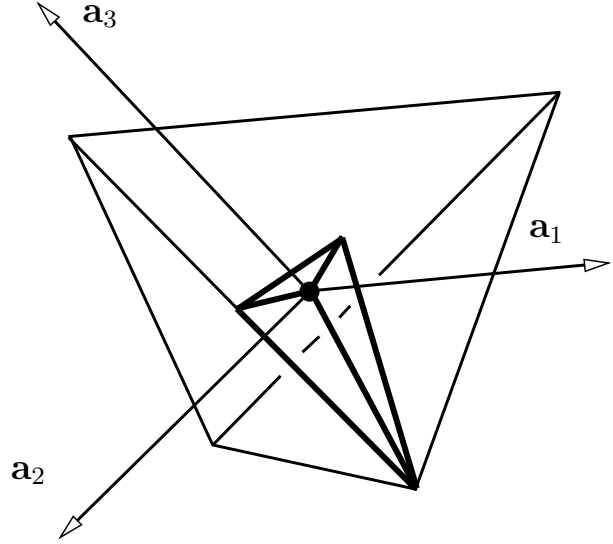} 
\end{center}
\caption{\label{fig:tetra1}The Weyl vectors $a_1,a_2,a_3$ of the Coxeter group $\Gamma=\circ  -\circ -\circ -\circ$, and the Coxeter simplex bounded by the Weyl reflection planes. $24$ Coxeter simplices share a vertex and   form the tetrahedral manifold $N1$. 
In Figs.~2-8 we replace the Platonic spherical polyhedra by their Euclidean counterparts.}
\end{figure}

The Coxeter group $\Gamma= \circ -\circ -\circ - \circ$ is isomorphic to the symmetric group $S(5)$ of order 
$|\Gamma|=5!=120$. On $S^3$ it has $120$ Coxeter simplices. Sets of $24$  of them, each sharing a single vertex, form 
$5$ tetrahedra, Fig.~\ref{fig:tetra1}. The four Weyl generators of $\Gamma=S(5)$ correspond to the four permutations $(1,2),(2,3),(3,4),(4,5)$
written in cycle form.

\begin{figure}[t]
\begin{center}
\includegraphics[width=0.5\textwidth]{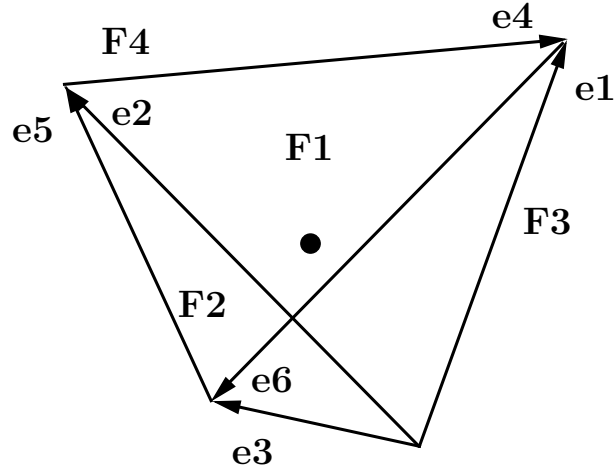} 
\end{center}
\caption{\label{fig:tetra2} Enumeration of the four faces Fs and six directed edges ej of the tetrahedral spherical manifold
from \cite{EV04}.}
\end{figure}

The tetrahedra  tile $S^3$ and form the $5$-cell tiling \cite{SO58} p. 170. 
In Fig.~\ref{fig:tetra2} we show the enumeration of faces and directed edges of the tetrahedron.

\paragraph{Face gluings.}
\begin{equation}
 \label{9b}
F3\cup F1,\: F2\cup F4.
\end{equation}
\paragraph{Edge gluing scheme.}
In this  and in the corresponding schemes for other manifolds, directed edges in a single horizontal line are glued.
\begin{equation}
 \label{9c}
\left[
\begin{array}{lll}
 1&\overline{3}&\overline{4}\\
2&\overline{5}&\overline{6}\\
\end{array} \right]
\end{equation}
The combination of  the given face and edge gluings fully determines the generators of the fundamental group. 

\subsubsection*{Group H=${\rm deck}(N1)$.}

The group $H={\rm deck}(N1)$ of deck transformations 
from \cite{KR08} is the cyclic group $C_5$. Its generator is given in  
Table~\ref{TableN1a}.

\begin{figure}[t]
\begin{center}
\includegraphics[width=0.5\textwidth]{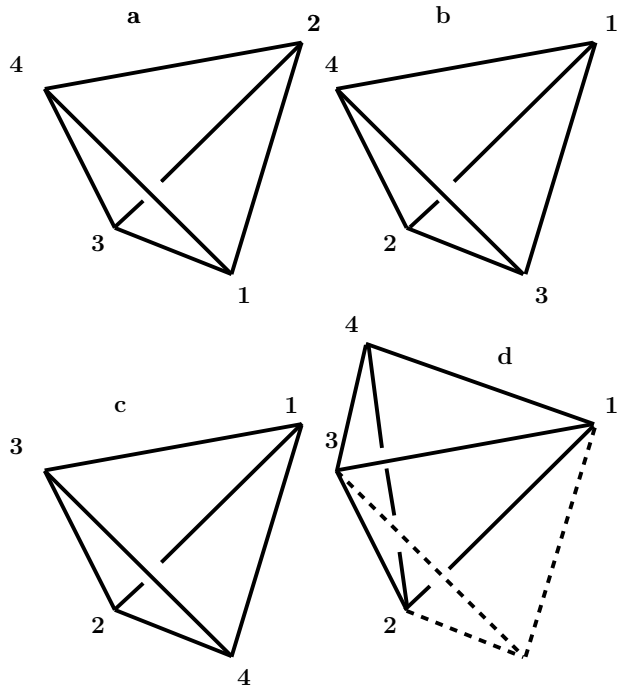} 
\end{center}
\caption{\label{fig:tetraglue} The action of the inverse generator $(W_4W_3W_2W_1)=(5,4,3,2,1)$ of $C_5$,
taken as a cyclic permutation from $\Gamma=S(5)$. The  vertices of the tetrahedral prototile  are  
denoted by $(1,2,3,4)$. Shown is the factorization of this generator into Weyl 
reflections.  a: initial tetrahedron $T$, 
b: $(W_2W_1)T$, c: $(W_3W_2W_1)T$, d: $(W_4W_3W_2W_1)T$. 
The reflection plane for $W_4$ contains  the vertices $(1,2,3)$ in c and d.
$W_4$ in d reflects the tetrahedron shown  in  c from the dashed into the undashed position.}
\end{figure}

For the generator of deck transformations of the tetrahedron we deviate from the gluing prescription of \cite{EV04}.
Instead of the generator $g_1$ from \cite{EV04} for the face gluing $F3\cup F1$ we prefer in Table~\ref{TableN1a}
the cyclic permutation 
$(1,2,3,4,5)$. The action of its inverse is illustrated in Fig.~\ref{fig:tetraglue}. It can be shown in terms of permutations in cycle form, that 
$g_1=(1,3,5,4,2)=(3,5,2)(1,2,3,4,5)(2,5,3)$ so that $g_1$ prescribed by \cite{EV04} is 
conjugate in $\Gamma$ to the present  choice. 

\begin{table}
\begin{eqnarray*}
\label{q1}
&&T_{(W_1W_2W_3W_4)}= T_{(g_l,g_r)},
\\ \nonumber
&&g_l=v_1v_2^{-1}v_3v_4^{-1}=
\left[
\begin{array}{ll}
\frac{2-2\sqrt{5}-i(\sqrt{2}+\sqrt{10})}{8}
&-\frac{-3\sqrt{2}+\sqrt{10}+i(6+2\sqrt{5})}{8\sqrt{3}}\\
\frac{-3\sqrt{2}+\sqrt{10}+i(-6-2\sqrt{5})}{8\sqrt{3}}
&\frac{2-2\sqrt{5} +i(\sqrt{2}+\sqrt{10})}{8} 
\end{array}
\right],
\\ \nonumber
&&g_r=v_1^{-1}v_2v_3^{-1}v_4= 
 \left[
\begin{array}{ll}
\frac{2+2\sqrt{5}+i(-\sqrt{2}+\sqrt{10})}{8}
&\frac{3\sqrt{2}+\sqrt{10}+i(-6+2\sqrt{5})}{8\sqrt{3}}\\
-\frac{3\sqrt{2}+\sqrt{10}+i(6-2\sqrt{5})}{8\sqrt{3}}
&\frac{2+2\sqrt{5} +i(\sqrt{2}-\sqrt{10})}{8} 
\end{array}
\right].
\end{eqnarray*}
\caption{\label{TableN1a} ($N1a$)
The generator of the cyclic group $H=C_5$ of deck transformations for the spherical tetrahedron. 
This generator corresponds to the product of the four generating Weyl reflections. The table is constructed by use of  eq.~\ref{8}.
}
\end{table}

The subgroups in $O(4,R)>S(5)>C_5$ and their reduction are implemented in \cite{KR08}.
In the next Table, a corrected version of Table 4.9 from \cite{KR08}, we give the multiplicity analysis with these subgroups
for $0\leq 2j \leq 10$.
The representations of $S(5)$ are characterized by partitions $f$. The Table shows that the  representations of $S(5)$ with partitions 
$f= \left[41\right],\:\left[2111\right]$ 
do not contribute $C_5$-invariant polynomials.
If random tetrahedral point symmetry is asssumed as suggested in \ref{sec:randompolyhedral} and 
carried out for the manifold $N2$ in \ref{sec:fromdeckvia}, 
one must look for polynomials invariant under $S\Gamma=A(5)$.  The corresponding  representations arise only from the partitions $\left[5\right]$ and 
$\left[11111\right]$ of $S(5)$. Table \ref{TableN1b} shows that then the total multiplicity of invariant polynomials under $A(5)$ for polynomial degrees $2j\leq 10$ reduces from $101$ to $13$.

\begin{table}
\begin{equation*}
\begin{array}{|r|rrrrrrrr|r|} \hline 
f:&& \left[5\right]&\left[1111\right]&\left[41\right]&\left[2111\right]&\left[32\right]
&\left[221\right]&\left[311\right]&m'((j,j),0)\\ 
\hline
&&&&&&&&&\\
 (2j)   &&   &  &  &  &  &  &  &\\
   
0   &&1 &  &  &  &  &  &  &1\\
1   &&  &  &1 &  &  &  &  &0\\
2   &&  &  &1 &  &1 &  &  &1\\
3   &&1 &  &1 &  &1 &  &1 &4\\
4   &&1 &  &2 &  &1 &1 &1 &5\\
5   &&1 &  &2 &  &2 &1 &2 &8\\
6   &&1 &  &3 &1 &3 &1 &2 &9\\
7   &&1 &  &4 &1 &3 &2 &3 &12\\
8   &&2 &  &4 &1 &4 &3 &4 &17\\
9   &&2 &  &5 &2 &5 &3 &5 &20\\
10  &&2 &1 &6 &2 &6 &4 &6 &25\\
\hline
&&&&&&&&&\\
\nu_0(f)&&12&1&0&0&26&15&48&102\\ \hline 
\end{array}
\end{equation*}
\caption{\label{TableN1b} ($N1b$)
Multiplicities $m((j,j),f)$ in the reduction of representations 
$D^{(j,j)}=\sum_j m((j,j),f)D^f$ from $O(4,R)$ to $S(5)$ as function of $(2j)=0,\ldots, 10$
and of all partitions $f$. 
$m'((j,j),0)$ in the last column denotes the total number of $C_5$-invariant modes 
for fixed $(2j)$, $\nu_0(f)$ in the last row those for a fixed partition $f$ up to $(2j)=10$.
}
\end{table}

\paragraph{Basis of harmonic analysis.}

Here we present  a new approach to the $C_5$-invariant basis by reducing directly between the groups 
$SO(4,R)>C_5$.
We first diagonalize the matrices $(g_l, g_r)$ eq.~\ref{q1} in the forms
\begin{equation}
 \label{t1}
g_l=c_l\delta_l c_l^{\dagger},\: g_r=c_r\delta_r c_r^{\dagger}.
\end{equation}
where the diagonal entries of $\delta_l, \delta_r$ are found from the traces of $(g_l,g_r)$ eq.~\ref{q1} as
$\lambda_l=\exp(\pm \frac{6i\pi}{10}),\:\lambda_r=\exp(\pm\frac{2i\pi}{10})$. 
Upon transforming  the coordinates $u$ from eq.~\ref{1}  by
\begin{equation}
 \label{t2}
u \rightarrow u'=c_l^{\dagger}uc_r,
\end{equation}
the Wigner polynomials $D^j(u')$ as functions of the new coordinates transform under $C_5$  by actions from left and right of 
diagonal $2 \times 2$ matrices as
\begin{equation}
\label{t3}
u' \rightarrow \delta_l^{-1} u'\delta_r.
\end{equation}
Under this substitution, the Wigner polynomials 
from eqs. \ref{a3}, \ref{t1} transform as 
\begin{equation}
 \label{t4}
D^j_{m_1m_2}(u') \rightarrow D^j_{m_1m_2}(\delta_l^{-1} u'\delta_r)= \exp(i(-3m_1+m_2)\frac{2\pi}{5}) D^j_{m_1m_2}(u').
\end{equation}
Projection to the identity representation of $C_5$ from this equation 
requires 
\begin{equation}
 \label{t5}
-3m_1+m_2\equiv 0\: {\rm mod}\: 5.
\end{equation}
This selection rule yields the basis of the harmonic analysis of the manifold $N1$ in Table~\ref{TableN1c}.

\begin{table}
\begin{equation*}
\begin{array}{|l|l|} \hline
&\\
\psi^j_{m_1m_2}(u'):& \delta_{-3m_1+m_2,\:0\: {\rm mod}\: 5}\: D^j_{m_1m_2}(u'),\: 2j=0,1,2,...\: -j\leq (m_1,m_2)\leq j.\\
&\\ \hline
\end{array}
\end{equation*}
\caption{\label{TableN1c} ($N1c$) The $C_5$-invariant basis of  harmonic analysis for the tetrahedral manifold $N1$ in terms
of Wigner polynomials.
}
\end{table}

\subsection{The cubic  manifolds $N2$ and $N3$.}
\label{subsec:thecubic}

\begin{figure}[tb]
\begin{center}
\includegraphics[width=0.5\textwidth]{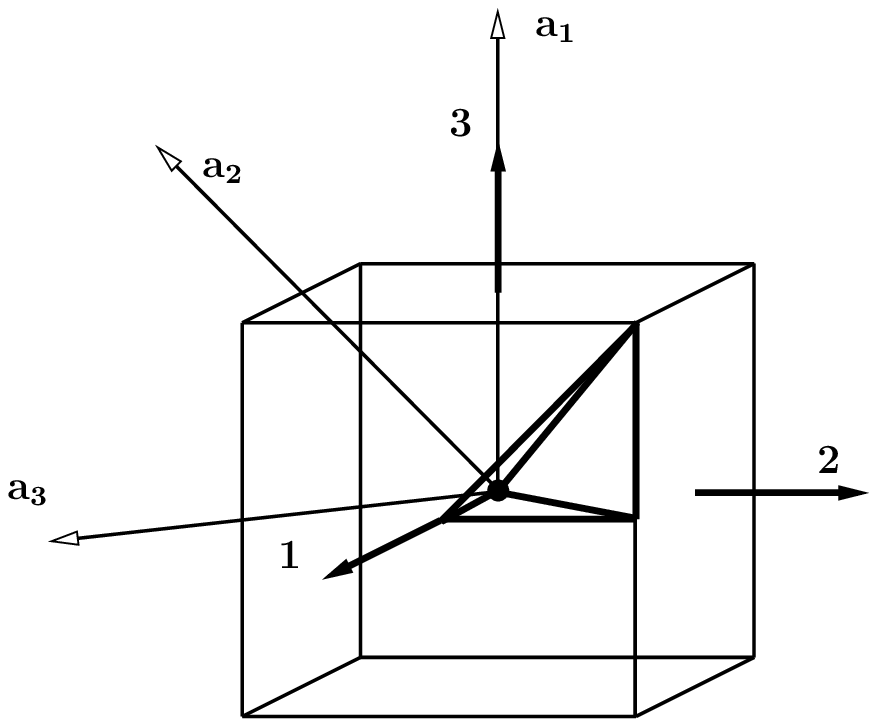} 
\end{center}
\caption{\label{fig:cubus1} The unit vectors $1,2,3$, the Weyl vectors $a_1,a_2,a_3$ of the Coxeter group $\Gamma=\circ \stackrel{4}{-} \circ -\circ -\circ$, and the Coxeter simplex bounded by the Weyl reflection planes. $48$ Coxeter simplices share a vertex and   form the cubic manifolds $N2, N3$.}
\end{figure}

\begin{figure}[tb]
\begin{center}
\includegraphics[width=0.5\textwidth]{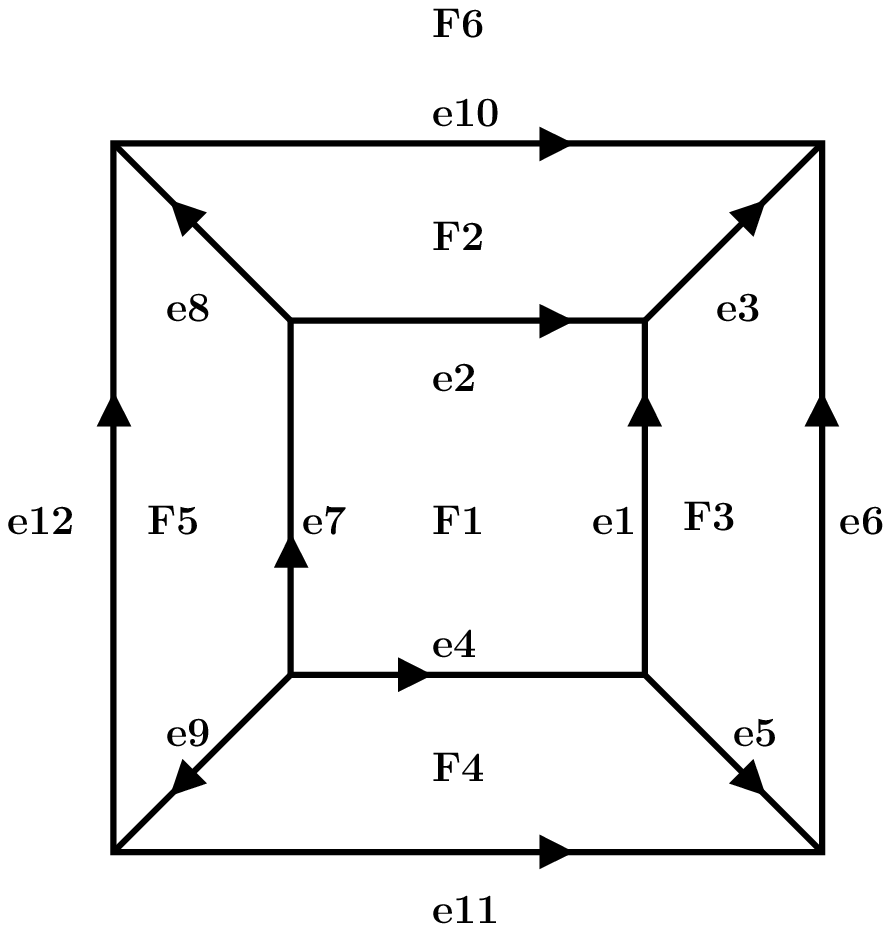} 
\end{center}
\caption{\label{fig:cubus2}
Enumeration of faces $F1,\ldots, F6$ and edges $e1,\ldots, e12$ for the
cubic prototile according to Everitt \cite{EV04} p.~260 Fig.~2.}
\end{figure}

The Coxeter group $\Gamma=\circ \stackrel{4}{-} \circ -\circ -\circ$ has $|\Gamma|=384$ simplices on $S^3$. Sets of $48$ of them sharing a single vertex form the $8$ cubes, Fig.~\ref{fig:cubus1}, of the $8$-cell tiling \cite{SO58} pp.~170-171. In Fig.~\ref{fig:cubus2} we show the enumeration of faces and edges of the cube. The generated  groups of deck transformations from \cite{KR09} are a cyclic group $H=C_8$ for $N2$ and the quaternion group $Q$ for $N3$.

\subsubsection*{Cubic manifold $N2$.}

\paragraph{Face gluings.}
After correction of  an  error in \cite{KR09} eq. 9,
\begin{equation}
\label{c9}
F3\cup F1,\; F4\cup F2,\; F6\cup F5.
\end{equation} 
\paragraph{Edge gluing scheme.}
Directed edges in a single line are glued.
\begin{equation}
\left[
\begin{array}{lll}
 1&3&4\\
2&6&\overline{9}\\
5&7&\overline{10}\\
8&11&\overline{12}\\
\end{array} \right]
\end{equation}

\subsubsection*{Group $H={\rm deck}(N2)$.}

The group $H={\rm deck}(N2)$ is the cyclic group $C_8$ generated by the elements in Table~\ref{TableN2a}.

\begin{table}
\begin{equation*}
\begin{array}{|l|l|ll|} \hline
t& (g_1)^t\;x &(g_l)^t&(g_r)^t\\
\hline
1& (x_1,-x_3,x_0,x_2)&
\left[\begin{array}{ll}
-\overline{a}&0\\
0&-a\\
\end{array}
\right]
&
\left[\begin{array}{ll}
0&-a^3\\
-a&0\\
\end{array}
\right]\\ \hline
2& (-x_3,-x_2,x_1,x_0)&
\left[\begin{array}{ll}
\overline{a}^2&0\\
0&a^2\\
\end{array}
\right]
&
\left[\begin{array}{ll}
-1&0\\
0&-1\\
\end{array}
\right]\\ \hline
3& (-x_2,-x_0,-x_3,x_1)&
\left[\begin{array}{ll}
-\overline{a}^3&0\\
0&-a^3\\
\end{array}
\right]
&
\left[\begin{array}{ll}
0&a^3\\
a&0\\
\end{array}
\right]\\ \hline
4& (-x_0,-x_1,-x_2,-x_3)&
\left[\begin{array}{ll}
-1&0\\
0&-1\\
\end{array}
\right]
&
\left[\begin{array}{ll}
1&0\\
0&1\\
\end{array}
\right]\\ \hline
5& (-x_1,x_3,-x_0,-x_2)&
\left[\begin{array}{ll}
\overline{a}&0\\
0&a\\
\end{array}
\right]
&
\left[\begin{array}{ll}
0&-a^3\\
-a&0\\
\end{array}
\right]\\ \hline
6& (x_3,x_2,-x_1,-x_0)&
\left[\begin{array}{ll}
-\overline{a}^2&0\\
0&-a^2\\
\end{array}
\right]
&
\left[\begin{array}{ll}
-1&0\\
0&-1\\
\end{array}
\right]\\ \hline
7& (x_2,x_0,x_3,-x_1)&
\left[\begin{array}{ll}
\overline{a}^3&0\\
0&a^3\\
\end{array}
\right]
&
\left[\begin{array}{ll}
0&a^3\\
a&0\\
\end{array}
\right]\\ \hline
8& (x_0,x_1,x_2,x_3)&
\left[\begin{array}{ll}
1&0\\
0&1\\
\end{array}
\right]
&
\left[\begin{array}{ll}
1&0\\
0&1\\
\end{array}
\right]\\ \hline
\end{array}
\end{equation*}
\caption{\label{TableN2a} ($N2a$)
The elements of the cyclic group $H={\rm deck}(N2)=C_8$ of deck transformations  of
the manifold $N2$ and their actions on $S^3$, with $a=\exp(\pi i/4)$.
}
\end{table}

The projector eq.~\ref{5b} for the manifold $N2$ is given in \cite{KR09}.

\paragraph{Basis of harmonic analysis:} Given in Table~\ref{TableN2b}. 

\begin{table}
\begin{equation*}
 \label{cc4}
\begin{array}{|l|} \hline
j={\rm integer},\, m_1={\rm even},\,-j\leq m_1\leq j,\, i^{m_1}(-1)^j=1, \, m_2=0:
\\ \hline
\\
\phi^j_{m_1,0}=\frac{\sqrt{2j+1}}{\sqrt{8}\pi}D^j_{m_1,0}(u),
\\  
\\ \hline
j={\rm integer},\, m_1={\rm even},\,-j\leq m_1\leq j,\, 0<m_2 \leq j:
\\ \hline
\\
\phi^j_{m_1,m_2}=\frac{\sqrt{2j+1}}{4\pi}
\left[D^j_{m_1,m_2}(u)+i^{m_1}(-1)^{(j+m_2)}i^{m_2}D^j_{m_1,-m_2}(u)\right] 
\\
\\ \hline
\end{array} 
\end{equation*}
\caption{\label{TableN2b} ($N2b$)
The $H=C_8$-periodic basis $\{\phi^j_{m_1,m_2}\}$ on $S^3$ for the harmonic analysis on
the cubic spherical 3-manifold $N2$ in terms of Wigner polynomials $D^j(u)$ on $S^3$.
}
\end{table}

\subsubsection*{Cubic manifold $N3$.}

\paragraph{Face gluings.} 
Opposite faces of the cube are glued,
\begin{equation}
\label{c34} 
F1\cup F6,\, F2\cup F4,\: F3\cup F5.
\end{equation}
\paragraph{Edge gluing scheme.}
Directed edges in a single line are glued.
\begin{equation}
 \label{c9b}
\left[
\begin{array}{lll}
 1&8&11\\
2&\overline{6}&\overline{9}\\
3&4&\overline{12}\\
5&\overline{7}&\overline{10}\\
\end{array} \right]
\end{equation}

\subsubsection*{Group $H={\rm deck}(N3)$}

We construct three glue generators $q_1, q_2, q_3$ in Table~\ref{TableN3a} from the prescription of 
\cite{EV04} p.~259 Table 3. 
The group $H$ is the quaternionic group $Q$ \cite{CO65} p. 134. It acts exclusively by left action.

\begin{table}
\begin{equation*}
\begin{array}{|l|l|l|l|} \hline
\label{c39}
i& q_i \: x&g_{li}&g_{ri}\\
\hline
1& (x_1,-x_0,x_3,-x_2)&
\left[
\begin{array}{ll}
0&-i\\
-i&0\\
\end{array}
\right]=-\mathbf{k}&e\\ \hline
2& (x_2,-x_3,-x_0,x_1)&
\left[
\begin{array}{ll}
0&-1\\
1&0\\
\end{array}
\right]=-\mathbf{j}&e\\ \hline
3& (x_3,x_2,-x_1,-x_0)&
\left[
\begin{array}{ll}
-i&0\\
0&i\\
\end{array}
\right]=-\mathbf{i}&e\\ \hline
\end{array}
\end{equation*}
\caption{\label{TableN3a} ($N3a$)
The three generators $q_i$ of the quaternionic group $H={\rm deck}(N3)=Q$  as elements
of the Coxeter group $\Gamma$ and the corresponding  pairs 
$(g_{li},g_{ri})\in (SU^l(2,R)\times SU^r(2,R))$. Products of the matrices $\mathbf{i}$, 
$\mathbf{j}$, $\mathbf{k}$ follow the standard quaternionic rules.
}
\end{table}

The projector eq.~\ref{5b} acting on Wigner polynomials from \cite{KR09} gives
\begin{equation}
\label{s6b}
(P^0_{Q}D^j_{m_1,m_2})(u)= \frac{1}{8}\left[1+(-1)^{2j}\right] \left[1+(-1)^{m_1}\right] \left[D^j_{m_1,m_2}(u)+(-1)^jD^j_{-m1,m_2}(u)\right].
\end{equation}

\paragraph{Basis of harmonic analysis:} Given in Table~\ref{TableN3b}.

\begin{table}
\begin{equation*}
\begin{array}{|l|} \hline
j={\rm odd}, j\geq 3,\, m_1={\rm even},\, 0<m_1\leq j,\, -j\leq m_2\leq j:
\\ \hline
\\
 \phi^{j {\rm odd}}_{m_1,m_2}=\frac{\sqrt{2j+1}}{4\pi}\left[ D^j_{m_1,m_2}(u)-D^j_{-m_1,m_2}(u)\right],
\\ 
 m(Q (j,j),0)=\frac{1}{2}(2j+1)(j-1),
\\
\\ \hline 
j={\rm even},\, m_1=0,\,  -j\leq m_2\leq j:
\\ \hline
\\
 \phi^{j {\rm even}}_{0,m_2}=\frac{\sqrt{2j+1}}{\sqrt{8}\pi}D^j_{0,m_2}(u),
\\
\\ \hline
j\geq 2,{\rm even},\,0<m_1\leq j,\,m_1={\rm even}:
\\ \hline
\\
 \phi^{j {\rm even}}_{m_1,m_2}=\frac{\sqrt{2j+1}}{4\pi}\left[ D^j_{m_1,m_2}(u)+D^j_{-m_1,m_2}(u)\right],
\\ 
m(Q (j,j),0)= \frac{1}{2}(2j+1)(j+2)
\\
\\ \hline 
\end{array} 
\end{equation*}
\caption{\label{TableN3b} ($N3b$)
The $Q$-invariant orthonormal basis 
$\{ \phi^{j {\rm odd}}_{m_1,m_2}, \phi^{j {\rm even}}_{m_1,m_2}\}$ for the harmonic analysis on the cubic spherical manifold $N3$ in terms of Wigner polynomials $D^j(u)$ on $S^3$.
}
\end{table}

\subsection{The octahedral manifolds $N4$, $N5$, $N6$.}

\begin{figure}[t]
\begin{center}
\includegraphics[width=0.5\textwidth]{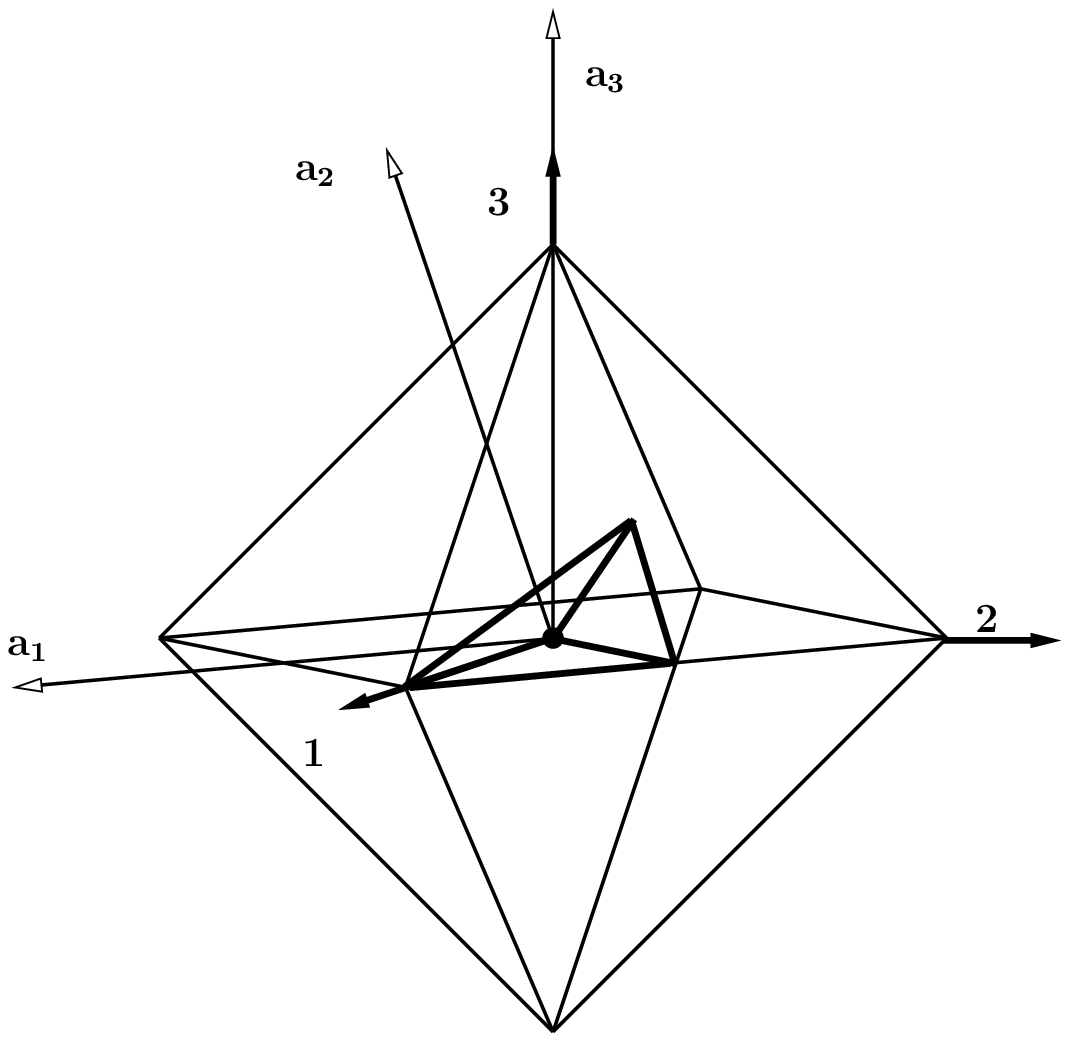}
\end{center}
\caption{\label{fig:octa1} The unit vectors $1,2,3$, the Weyl vectors $a_1, a_2, a_3$ of the Coxeter group $\Gamma= \circ -\circ \stackrel{4}{-}\circ - \circ$, and the Coxeter simplex bounded by the Weyl reflection planes.  $48$ Coxeter simplices share a vertex and 
form the octahedral manifolds $N4, N5, N6$.}
\end{figure}

\begin{figure}[t]
\begin{center}
\includegraphics[width=0.5\textwidth]{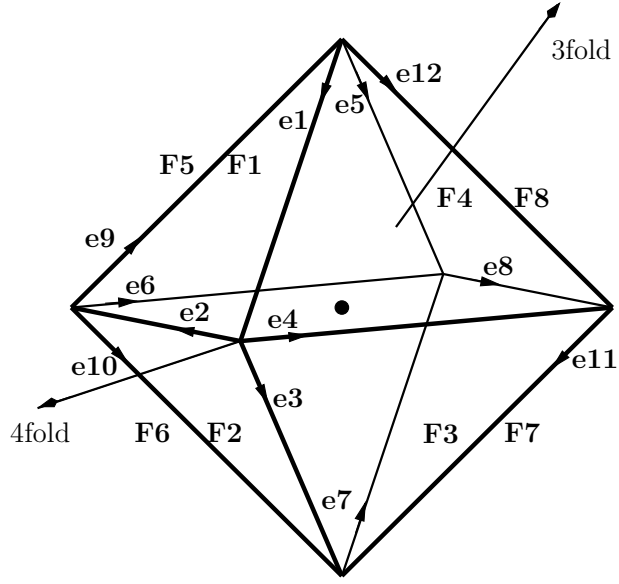}
\end{center}
\caption{\label{fig:octa2} The octahedron projected to the plane with faces $F1\ldots F8$ and directed edges $e1\ldots e12$
 according to \cite{EV04}. The products of Weyl reflections $(W_1W_2)$ and $(W_2W_3)$ generate right-handed
$3$fold and $4$fold rotations respectively.}
\end{figure}

The Coxeter group $\Gamma=\circ -\circ\stackrel{4}{-} \circ -\circ$ has $|\Gamma|=1152$ simplices on $S^3$. Sets of $48$ of them sharing a single vertex form the $24$ octahedra, Fig.~\ref{fig:octa1}, of the $24$-cell tiling \cite{SO58} pp.~171-172. The homotopies of the octahedral 3-manifolds 
given in \cite{EV04} were corrected in part in \cite{CAV09}. Face and edge enumerations are given in Fig.~\ref{fig:octa2}.

The groups of deck transformations from \cite{KR09b} are the direct product $H= C_3^l \times Q^r$ for 
N4, a group $H=B$  for N5, and the binary tetrahedral group ${\cal T}^*$ for N6. The generators of these groups are 
given in Tables~\ref{TableN4a}, \ref{TableN4b}, \ref{TableN5a}.

\subsubsection*{Octahedral manifold $N4$.}

\paragraph{Face gluings:} 

\begin{equation}
F6\cup F2, F5\cup F3, F1\cup F4,\: F7\cup F8.
\end{equation}

\paragraph{Edge gluing scheme:}

\begin{equation}
\label{oc9}
\left[
\begin{array}{lll}
1&4&9\\
2&7&\overline{12}\\
3&6&\overline{10}\\
5&8&11\\
\end{array}
\right]
\end{equation}

\subsubsection*{Group $H={\rm deck}(N4)$.}

The group $H={\rm deck}(N4)$ is a direct product $H=C_3^l \times Q^r$ where the upper indices stand for left and right actions.
For the projection to a $H$-periodic basis of $N4$ we  first diagonalize the generator $-\alpha_2\in C_3^l$
given in Table~\ref{TableN6b},
\begin{eqnarray}
\label{oc11d}
&& -\alpha_2= c 
\left[
\begin{array}{ll}
 \exp(\frac{2\pi i}{3})&0\\
0&\exp(-\frac{2\pi i}{3})
\end{array} \right]
c^{\dagger},
\\ \nonumber
&& c=
\left[
\begin{array}{ll}
(1-i)\frac{-1+\sqrt{3}}{2\sqrt{3-\sqrt{3}}}& -(1-i)\frac{1+\sqrt{3}}{2\sqrt{3+\sqrt{3}}}\\
\frac{1}{\sqrt{3-\sqrt{3}}}& \frac{1}{\sqrt{3+\sqrt{3}}}
\end{array} 
\right].
\end{eqnarray}
Similar as was done for the tetrahedral manifold, 
we interprete the substitution  $u \rightarrow u'=c^{\dagger}u$
as a transformation to new coordinates $u'$ and derive the basis in these new coordinates. All elements of the group $C_3^l$ are 
now diagonal in the new coordinate basis. Projection to the identity representation then gives the result of Table~\ref{TableN4c}. 

\begin{table}
\begin{equation*}
\begin{array}{|l|l|l|}
\hline
 g& g_l& g_r\\ 
\cline{1-3}
g_1 & -\alpha_2&\mu\\
\hline
g_2 & -\alpha_2^{-1}&-e\\
\hline 
g_3 &\alpha_2&\nu\\
\hline
g_4 &\alpha_2^{-1}&\omega\\ 
\hline
\end{array}
\end{equation*}
\caption{\label{TableN4a} ($N4a$)
Generators $g=(g_l,g_r)$ of $deck(N4)$.
We use the short-hand notation of Table~\ref{TableN6b}.
}
\end{table}

From these generators we derive the structure of the group $H=C_3^l \times Q^r$ with elements given in Table~\ref{TableN4b}.

\begin{table}
\begin{equation*}
\begin{array}{|l|l|} \hline 
{\rm subgroup}& {\rm elements}\: (g_l,g_r)\\ \hline
C_3^l& ( -\alpha_2,e),((\alpha_2)^2,e),( (-\alpha_2)^3,e)=(e,e) \\ \hline
Q^r& (e,\pm e), (e,\pm \mu), (e, \pm \nu), (e, \pm\omega)\\ \hline
\end{array} 
\end{equation*}
\caption{\label{TableN4b} ($N4b$)
The elements $g=(g_l, g_r)$ of the group ${\rm deck}(N4)= C_3^l\times Q^r$ in the notation of Table~\ref{TableN6b}.
}
\end{table}

\paragraph{Basis of harmonic analysis:} Given in Table~\ref{TableN4c}.

\begin{table}
\begin{equation*}
\begin{array}{|l|} \hline
j={\rm odd},\: j\geq 3,\: m_2={\rm even},\: 0<m_2 \leq j,\: m_1=\rho \equiv 0\: {\rm mod}\: 3:
\\ \hline
\\ 
 \phi^{j{\rm odd}}_{\rho,m_2}= \left[ D^j_{\rho,m_2}(u')-D^j_{\rho,-m_2}(u')\right],
\\
\\ \hline 
j={\rm even},\: m_2=0,\: m_1=\rho \equiv 0\: {\rm mod}\: 3:
\\  \hline
\\
\phi^{j{\rm even}}_{\rho,0}= D^j_{\rho, 0}(u') 
\\
\\ \hline
j \geq 2,\:   {\rm even},\: 0<m_2\leq j,\: m_2={\rm even},\: m_1=\rho \equiv 0\:  {\rm mod}\: 3:
\\ \hline 
\\
\phi^{j{\rm even}}_{\rho,m_2}= \left[D^j_{\rho, m_2}(u') +D^j_{\rho,-m_2}(u')\right] 
\\
\\ \hline
\end{array} 
\end{equation*}
\caption{\label{TableN4c} ($N4c$)
The $(C_3^l \times Q^r)$-periodic basis for the manifold $N4$ in terms of Wigner polynomials $D^j$.
Only integer values of $j$ appear. The coordinate transform $u \rightarrow u'=c^{\dagger}u$ in $D^j(u)$  follows with 
$c$ from eq.~\ref{oc11d}.
} 
\end{table}

\subsubsection*{Octahedral manifold $N5$.}

\paragraph{Face gluings:} 

\begin{equation}
F6 \cup F8, F1 \cup F4, F2 \cup F7, F3\cup F5.
\end{equation}

\paragraph{Edge gluing scheme:}

\begin{equation}
\label{oc12}
\left[
\begin{array}{lll}
1&4&9\\
2&\overline{7}&\overline{12}\\
3&6&8\\
5&\overline{10}&11\\
\end{array}
\right]
\end{equation}

\subsubsection*{Group $H={\rm deck}(N5)$.}

For this manifold we denote the group of deck transformations by $H={\rm deck}(N5)=:B$ and give its elements in Table~\ref{TableN5a}.

\begin{table}
\begin{equation*}
\begin{array}{|l|l|l|l|} \hline
s& g_l         &g_r     &g_l^{-1}g_r\\ \hline
\pm 1&\alpha_2^{-1}& \mp \nu&\pm \alpha_1\\ \hline
\pm 2&\alpha_2^{-1}&\pm e&\pm \alpha_2\\ \hline
\pm 3&\alpha_2     &\pm \nu &\pm \alpha_3\\ \hline
\pm 4&
\sqrt{\frac{1}{2}}
\left[
\begin{array}{ll}
-i&-1\\
1&i
\end{array}
\right]&
\pm \sqrt{\frac{1}{2}}
\left[
\begin{array}{ll}
1&-1\\
1&1
\end{array}
\right]&
\pm \alpha_4\\ \hline
\pm 5&
\sqrt{\frac{1}{2}}
\left[
\begin{array}{ll}
-i&-1\\
1&i
\end{array}
\right]&
\mp \sqrt{\frac{1}{2}}
\left[
\begin{array}{ll}
1&1\\
-1&1
\end{array}
\right]&
\pm \alpha_1^{-1}\\ \hline
\pm 6&\alpha_2&\pm e&\pm \alpha_2^{-1}\\ \hline
\pm 7&
\left[
\begin{array}{ll}
0&\overline{\theta}\\
-\theta&0
\end{array}
\right]&
\mp \sqrt{\frac{1}{2}}
\left[
\begin{array}{ll}
1&-1\\
1&1
\end{array}
\right]&
\pm \alpha_3^{-1}\\ \hline
\pm 8&
\left[
\begin{array}{ll}
0&\overline{\theta}\\
-\theta&0
\end{array}
\right]&
\pm \sqrt{\frac{1}{2}}
\left[
\begin{array}{ll}
1&1\\
-1&1
\end{array}
\right]&
\pm \alpha_4^{-1}\\ \hline
\pm 9&
e&\pm e&\pm e\\ \hline
\pm 10&
-\sqrt{\frac{1}{2}}
\left[
\begin{array}{ll}
i&i\\
i&-i
\end{array}
\right]&
\pm \sqrt{\frac{1}{2}}
\left[
\begin{array}{ll}
1&1\\
-1&1
\end{array}
\right]&
\pm \mu\\ \hline
\pm 11&e&\pm\nu&\pm \nu\\ \hline
\pm 12&
\sqrt{\frac{1}{2}}
\left[
\begin{array}{ll}
i&i\\
i&-i
\end{array}
\right]&
\pm \sqrt{\frac{1}{2}}
\left[
\begin{array}{ll}
1&-1\\
1&1
\end{array}
\right]&
\pm \omega\\ \hline 
\end{array}
\end{equation*}
\caption{\label{TableN5a} ($N5a$)
Elements $g_j=(g_l,g_r),\: s= \pm 1,..., \pm 12$ of the group $B={\rm deck}(N5)$, enumerated according to the $24$ octahedral center positions $u''=g_l^{-1}g_r\in S^3$,  in the order and notation of Table~\ref{TableN6b}.
}
\end{table}

\paragraph{Basis of harmonic analysis:}

The projection and multiplicity must be computed with eqs.~\ref{5b}, \ref{5c}.

\subsubsection*{Octahedral manifold $N6$.}

\paragraph{Face gluings:}

\begin{equation}
F6\cup F4, F5\cup F3, F8\cup F2, F7\cup F1.
\end{equation}

\paragraph{Edge gluing scheme:}

\begin{equation}
\label{oc15}
\left[
\begin{array}{lll}
1&8&10\\
2&5&11\\
3&6&12\\
4&7&9\\
\end{array}
\right]
\end{equation}

\subsubsection*{Group $H={\rm deck}(N6)$.}

The group $H={\rm deck}(N6)$ is the binary tetrahedral group ${\cal T}^*$.

\begin{table}
\begin{equation*}
\begin{array}{|l|l|l|}
\hline 
 g& g_l& g_r\\ 
\cline{1-3}
&&\\
g_1 & 
\sqrt{\frac{1}{2}}\left[
\begin{array}{ll}
\theta&\theta\\
-\overline{\theta}&\overline{\theta}
\end{array}
 \right]:=\alpha_1
&
e\\
\cline{1-3}
&&\\
g_2 & 
\sqrt{\frac{1}{2}}
\left[
\begin{array}{ll}
\overline{\theta}&\theta\\
-\overline{\theta}&\theta
\end{array}
 \right]:=\alpha_2^{-1}
&
e\\
\cline{1-3}
&&\\
g_3 & 
\sqrt{\frac{1}{2}}\left[
\begin{array}{ll}
\overline{\theta}&-\overline{\theta}\\
\theta&\theta
\end{array}
 \right]:= \alpha_4^{-1}
&
e\\
\cline{1-3}
&&\\
g_4 & 
\sqrt{\frac{1}{2}}\left[
\begin{array}{ll}
\theta&-\overline{\theta}\\
\theta&\overline{\theta}
\end{array}
 \right]:=\alpha_3
&
e\\
\hline
\end{array}
\end{equation*}
\caption{\label{TableN6a} ($N6a$)
Generators $g=(g_l,g_r)$ of $deck(N6)$, compare Table~\ref{TableN6b}.
}
\end{table}

Using the equivalence $(g_l,g_r)\sim (-g_l, -g_r)$, we have written  $H$ entirely in terms of left actions. The group $H$ of homotopies and deck transformations 
of the 3-manifold $N6$ then turns out to be the
binary tetrahedral group $<2,3,3>$ of order $24$ in the notation of Coxeter and Moser \cite{CO65} pp.~134-135. The elements and multiplication rules are 
given in Tables~\ref{TableN6b}, \ref{TableN6c}.

\begin{table}
\begin{equation}
\begin{array}{|l|l|l|l|} \hline
\alpha_1&\alpha_2&\alpha_3&\alpha_4\\ 
\cline{1-4}
&&&\\
\sqrt{\frac{1}{2}}
\left[\begin{array}{ll}
\theta &\theta\\
-\overline{\theta}&\overline{\theta}
\end{array}
\right]
&
\sqrt{\frac{1}{2}}
\left[\begin{array}{ll}
\theta &-\theta\\
\overline{\theta}&\overline{\theta}
\end{array}
\right]
& 
\sqrt{\frac{1}{2}}
\left[\begin{array}{ll}
\theta &-\overline{\theta}\\
\theta&\overline{\theta}
\end{array}
\right]
& 
\sqrt{\frac{1}{2}}
\left[\begin{array}{ll}
\theta &\overline{\theta}\\
-\theta&\overline{\theta}
\end{array}
\right]
\\
\cline{1-4}
&&&\\
\alpha_1^{-1}& \alpha_2^{-1}&\alpha_3^{-1}&\alpha_4^{-1}\\
\hline
&&&\\
\sqrt{\frac{1}{2}}
\left[\begin{array}{ll}
\overline{\theta} &-\theta\\
\overline{\theta}&\theta
\end{array}
\right]
& 
\sqrt{\frac{1}{2}}
\left[\begin{array}{ll}
\overline{\theta} &\theta\\
-\overline{\theta}&\theta
\end{array}
\right]
& 
\sqrt{\frac{1}{2}}
\left[\begin{array}{ll}
\overline{\theta} &\overline{\theta}\\
-\theta&\theta
\end{array}
\right]
& 
\sqrt{\frac{1}{2}}
\left[\begin{array}{ll}
\overline{\theta} &-\overline{\theta}\\
\theta&\theta
\end{array}
\right]
\\
\cline{1-4}
&&&\\
e,-e&\mu&\nu& \omega\\
\hline
&&&\\
\left[\begin{array}{ll}
1&0\\
0&1
\end{array}
\right],
-\left[\begin{array}{ll}
1&0\\
0&1
\end{array}
\right]
&
\left[\begin{array}{ll}
0&i\\
i&0
\end{array}
\right]
&
\left[\begin{array}{ll}
0&-1\\
1&0
\end{array}
\right]
&
\left[\begin{array}{ll}
-i&0\\
0&i
\end{array}
\right]
\\
\hline
&&&\\
e^{-1}=e,(-e)^{-1}=-e&\mu^{-1}=-\mu&\nu^{-1}=-\nu& \omega^{-1}=-\omega\\
\hline
\end{array}
\end{equation}
\caption{\label{TableN6b} $(N6b)$
The binary tetrahedral group ${\cal T}^* \sim {\rm deck}(N6)$ has $16$ elements $\pm \alpha_j, \pm \alpha_j^{-1}$ and
$8$ elements $\pm e, \pm \mu, \pm \nu,\pm \omega$, with $\theta=\exp(i\pi/4),\: \overline{\theta}=\exp(-i\pi/4)$.
It acts from the left on $u \in S^3$.
}
\end{table}

The elements in Table~\ref{TableN6b} obey 
\begin{eqnarray}
&&(\alpha_j)^3=(\alpha_j)^{-3}=-e,\: \frac{1}{2}Tr(\alpha_j)=\frac{1}{2}Tr(\alpha_j^{-1})=
\frac{1}{2},\: j=1,..,4.
\\ \nonumber 
&& \mu^2=\nu^2=\omega^2=-e
\end{eqnarray}
The last four elements generate as subgroup the quaternion group $Q$, \cite{CO65}, pp.~134-135.
of order $8$ with standard elements $\mathbf{i}=-\omega,\:\mathbf{j}=-\nu,\:\mathbf{k}=\mu$.

\begin{table}
\begin{equation}
 \begin{array}{|l|llll|llll|llll|} \hline
&\alpha_1&\alpha_2&\alpha_3&\alpha_4&
\alpha_1^{-1}&\alpha_2^{-1}&\alpha_3^{-1}&\alpha_4^{-1}&
\mu&\nu&\omega&e\\ \hline
\alpha_1&
-\alpha_1^{-1}&\alpha_4&-\omega&-\nu&
e&\mu&\alpha_2^{-1}&\alpha_3&
-\alpha_3^{-1}&\alpha_2&\alpha_4^{-1}&\alpha_1\\
\alpha_2&
\alpha_3&-\alpha_2^{-1}&\nu&-\omega&
-\mu&e&\alpha_4&\alpha_1^{-1}&
\alpha_4^{-1}&-\alpha_1&\alpha_3^{-1}&\alpha_2\\
\alpha_3&
\mu&-\omega&-\alpha_3^{-1}&\alpha_1&
\alpha_2&\alpha_4^{-1}&e&\nu&
-\alpha_4&-\alpha_2^{-1}&\alpha_1^{-1}&\alpha_3\\
\alpha_4&
-\omega&-\mu&\alpha_2&-\alpha_4^{-1}&
\alpha_3^{-1}&\alpha_1&-\nu&e&
\alpha_3&\alpha_1^{-1}&\alpha_2^{-1}&\alpha_4\\ \hline
\alpha_1^{-1}&
e&\nu&\alpha_4^{-1}&\alpha_2&
-\alpha_1&\alpha_3^{-1}&-\mu&\omega&
\alpha_2^{-1}&-\alpha_4&-\alpha_3&\alpha_1^{-1}\\
\alpha_2^{-1}&
-\nu&e&\alpha_1&\alpha_3^{-1}&
\alpha_4^{-1}&-\alpha_2&\omega&\mu&
-\alpha_1^{-1}&\alpha_3&-\alpha_4&\alpha_2^{-1}\\
\alpha_3^{-1}&
\alpha_4&\alpha_1^{-1}&e&-\mu&
\omega&-\nu&-\alpha_3&\alpha_2^{-1}&
\alpha_1&\alpha_4^{-1}&-\alpha_2&\alpha_3^{-1}\\
\alpha_4^{-1}&
\alpha_2^{-1}&\alpha_3&\mu&e&
\nu&\omega&\alpha_1^{-1}&-\alpha_4&
-\alpha_2&-\alpha_3^{-1}&-\alpha_1&\alpha_4^{-1}\\ \hline
\mu&
-\alpha_2&\alpha_1&-\alpha_1^{-1}&\alpha_2^{-1}&
\alpha_3&-\alpha_4&\alpha_4^{-1}&-\alpha_3^{-1}&
-e&-\omega&\nu&\mu\\
\nu&
\alpha_4^{-1}&-\alpha_3^{-1}&-\alpha_4&\alpha_3&
-\alpha_2^{-1}&\alpha_1^{-1}&\alpha_2&-\alpha_1&
\omega&-e&-\mu&\nu\\
\omega&
\alpha_3^{-1}&\alpha_4^{-1}&\alpha_2^{-1}&\alpha_1^{-1}&
-\alpha_4&-\alpha_3&-\alpha_1&-\alpha_2&
-\nu&\mu&-e&\omega\\
e&\alpha_1&\alpha_2&\alpha_3&\alpha_4&
\alpha_1^{-1}&\alpha_2^{-1}&\alpha_3^{-1}&\alpha_4^{-1}&
\mu&\nu&\omega&e\\ \hline
\end{array}
\end{equation}
\caption{\label{TableN6c} ($N6c$)
Multiplication table for $12$ elements $g$ of the binary tetrahedral group ${\rm deck}(N6)$ given in Table~\ref{TableN6b}. The $12$ elements $-g$ have been 
suppressed.
}
\end{table}

\paragraph{Basis of harmonic analysis.}

The projection and multiplicity must be computed with eqs. \ref{5b}, \ref{5c}.

\subsection{The dodecahedral manifold N1'.}
This is the Poincar\'e dodecahedral manifold analyzed in \cite{KR05}. The Coxeter group $\circ -\circ -\circ \stackrel{5}{-} \circ$ on $S^3$ has 
$|\Gamma|= (120)^2$  simplices. The tiling on  $S^3$ is the $120$-cell \cite{SO58} pp.~176-177. 

The face gluings for this manifold are well known, see \cite{SE34} pp.~214-218.

\subsubsection*{Group $H={\rm deck}(N1')$.}

The homotopy group $\pi_1(N1')$ is the binary icosahedral  group ${\cal J}_2$ discussed in detail in \cite{KL84}.
In \cite{KR05} it is transformed into the isomorphic group $H={\rm deck}(N1')$ and related to
Hamilton's icosians.

\paragraph{Basis of harmonic analysis:}

The polynomial basis of the harmonic analysis on this manifold can be constructed for each degree $2j$ by the diagonalization of
an operator with explicit matrix representation given in \cite{KR05}, eq.(47) and Appendix. The multiplicity $m(j,0)$ of 
${\cal J}_2$-invariant 
basis functions is given from character analysis eq.~\ref{5c}, compare \cite{KR06}, by 

(i) the starting values
\begin{eqnarray}
\label{j1}
&&j\leq 30:\: m(j,0)=1\: {\rm for} 
\\ \nonumber
&&j=0,6,10,12,15,16,18,20,21,22,24,25,26,27,28,
\\ \nonumber
&& m(j,0)=0\: {\rm otherwise},
\end{eqnarray}
(ii) the recursion relation 
\begin{equation}
 \label{j2}
m(j+30,0)=m(j,0)\:+1.
\end{equation}

\section{Wigner polynomials.}
\label{sec:Wigner}

The Wigner polynomials are the spherical harmonics on the coset space\\ $SO(4,R)/SU^C(2,C)\sim SU^r(2,C)\sim S^3$,
see eq.~\ref{4}. From  \cite{WI59} pp.~163-166 they are 
homogeneous of degree $2j$ and   given in terms of the complex matrix elements of $u$ from eq.~\ref{1} by 
\begin{eqnarray}
 \label{A1}
&&D^j_{m_1m_2}(z_1,z_2,\overline{z}_1,\overline{z}_2)
=\left[\frac{(j+m_1)!(j-m_1)!}{(j+m_2)!(j-m_2)!}\right]^{1/2}
\\ \nonumber
&&\times \sum_{\sigma}\frac{(j+m_2)!(j-m_2)!(-1)^{m_2-m_1+\sigma}}
{(j+m_1-\sigma)!(m_2-m_1+\sigma)!\sigma!(j-m_2-\sigma)!}
\\ \nonumber
&&\times z_1^{j+m_1-\sigma}\overline{z}_2^{m_2-m_1+\sigma}z_2^{\sigma}\overline{z}_1^{j-m_2-\sigma},
\\ \nonumber 
&&2j=0,1,2,...,\: -j\leq (m_1,m_2)\leq j.
\end{eqnarray}
The summation over $\sigma$ is restricted by the inverse factorials.
The symmetries under inversion and complex conjugation of $u$ are 
\begin{equation}
 \label{A2}
D^j_{m_1m_2}(u^{-1})=\overline{D^j_{m_2m_1}(u)},\:\: \overline{D^j_{m_1m_2}(u)}=D^j_{m_1m_2}(\overline{u})
\end{equation}
In \cite{KR05} p. 3526 Lemma 5 it is shown that under  the Laplacian $\Delta$ on $E^4$ one has 
\begin{equation}
\label{A3}
\Delta D^j_{m_1m_2}(u)= (\sum_{i=0}^3 \frac{\partial^2}{\partial x_i^2})D^j_{m_1m_2}(u)=0. 
\end{equation}
In other words  the Wigner polynomials are harmonic.
For the Euler angle parametrization, orthogonality and completeness of the $D^j$ on $S^3 \sim SU(2,C)$ we refer to \cite{WI59}. 
The measure of integration on $S^3$ in terms of the Euler angles is, \cite{ED57} pp.~62-64,
\begin{equation}
 \label{A4}
d\mu(\alpha,\beta,\gamma)= d\alpha \sin(\beta) d\beta d\gamma,
\: \int_{SU(2,C)}d\mu(\alpha,\beta,\gamma)= 8\pi^2. 
\end{equation}

\section{From deck via point to unimodular  invariance.}\label{sec:fromdeckvia}

When we introduced in section~\ref{sec:randompolyhedral} the point group M of a Platonic manifold ${\cal M}$, 
we did not discuss its relation to the group ${\rm deck}({\cal M})$. 

In general, the group acts fixpoint-free on $S^3$ whereas M fixes the center of the prototile.
It follows that the two groups have the intersection $deck({\cal M})\cap M=e$. Both groups are subgroups of the 
Coxeter group, and so  products of their elements must generate a subgroup of $\Gamma$.
We place the centers of all  prototiles at $x=(1,0,0,0)$.
Then the action of a binary point group $M^*$ as a subgroup
of the diagonal group $SU^C(2,C)$ with elements of the form $g= (h,h)$, reduces
to the ordinary action $R(h)(x_1,x_2,x_3)$ with $R(h)\in M$.

The groups generated from deck and point groups and their projectors can be constructed for each manifold.
We exemplify the construction  by the cubic spherical manifold $N3$, with H the quaternion group $Q, |Q|=8$, and M the cubic point group $O, |O|=24$. The corresponding binary cubic group we denote by $O^*$.

By explicit computation one finds:

\paragraph{Prop C1:} Under conjugation with elements $h \in O^*$, the quaternion group Q is transformed into itself,
\begin{equation}
 \label{B1}
h \in O^*:\: h^{-1}Qh=Q.
\end{equation}
This implies that the group generated from both Q and $O$  is a semidirect group, with $Q$ a normal subgroup.

\paragraph{Prop C2:} For the manifold $N3$, the group generated by both H=Q and M=O is the semidirect group
\begin{equation}
 \label{B2}
G= Q \times_s O,\: |  Q \times_s O |=8 \times 24=192.
\end{equation}
with elements the products $ g= q\: h,\: q\in Q,\: h \in O^*$. 

This is also the  order of the unimodular subgroup 
$S\Gamma$ of 
the Coxeter group\newline
  $\Gamma=\circ \stackrel{4}{-} \circ -\circ -\circ $ for cubic 3-manifolds from Table~\ref{Table1}.
It is easy to show that the group $Q \times_s O$  exhausts and so is identical to 
this subgroup. The unimodular group $S\Gamma=Q\times_s O$ contains the two alternative deck groups 
for the cubic 3-manifolds $N2, N3$, and their cubic point symmetry group $O$.

We look for the projector to the identity representation of the group $Q \times_s O$.
From the semidirect product form eq.~\ref{B2}  there  follows:

\paragraph{Prop C3:} The projector to the identity representation of $Q \times_s O$ factorizes as
\begin{eqnarray}
\label{B3}
&& P^0_{Q \times_s O}= \frac{1}{|Q|\: |O|}\sum_{(q_rh_s,h_s) \in G} T_{(q_rh_s, h_s)}=P^0_{Q}\times P^{\Gamma_1},
\\ \nonumber
&&P^{\Gamma_1}= \frac{1}{|O|} \sum_{h \in O^*} T_{(h,h)}.
\end{eqnarray}
into the projectors of its two subgroups, with the quaternionic projector given in eq.~\ref{s6b}. Here  the sum over $h\in O^*$ can be restricted to the 24 elements of $O$.

We now construct the onset polynomial for the cubic spherical manifold $N3$ under $O$. From Table~\ref{Table61} it has  $j=2,\:l=4$. If we go to the alternative basis
eq.~\ref{a4}, we can use the  classical lowest cubic spherical harmonic, given in  \cite{KN64} pp.~108-109:
\begin{equation}
\label{B4}
\psi^{\Gamma_1}= \sqrt{\frac{7}{12}}Y^4_0+\sqrt{\frac{5}{24}}(Y^4_4+Y^4_{-4}).  
\end{equation}
Upon  using the same  linear $m$-combination in  the basis eq.~\ref{a4}, we pass to Wigner polynomials and find
\begin{eqnarray}
\label{B5}
&&\psi^{\Gamma_1}=\sqrt{\frac{7}{12}}\left[D^2_{2,2}(u)\langle 2-222|40\rangle + D^2_{-2,-2}(u)\langle 222-2|40\rangle\right]
\\ \nonumber
&&+\sqrt{\frac{7}{12}}\left[D^2_{1,1}(u)\langle 2-121|40\rangle(-1) +D^2_{-1,-1}(u)\langle 212-1|40\rangle(-1)
+D^2_{0,0}(u)\langle 2020|40\rangle\right]
\\ \nonumber
&&+\sqrt{\frac{5}{24}}\left[D^2_{-2,2}(u)\langle 2222|44\rangle +D^2_{2,-2}(u)\langle 2-22-2|4-4\rangle\right]
\end{eqnarray}
The relevant Wigner coefficients can be  found from \cite{ED57} p.~45:
\begin{equation}
\label{B5b}
\langle 2\mp 22\pm 2|40\rangle=\sqrt{\frac{1}{8!}}\: 4!,\:
\langle 2020|40\rangle=\sqrt{\frac{1}{8!}}\: 4!\: 6,\:\langle 2\pm 22\pm 2|4\pm 4\rangle=1.
\end{equation}
Next we follow eq.~\ref{a4}, apply the projector eq.~\ref{s6b} of the quaternion group $Q$ to the polynomial eq.~\ref{B5}, 
and obtain 
\begin{eqnarray}
 \label{B6}
&&\psi^{0,\Gamma_1}=P^0_{Q}\psi^{\Gamma_1}=
\sqrt{\frac{7}{12}}\left[\frac{1}{2}(D^2_{2,2}(u)+D^2_{-2,2}(u))\langle 2-222|40\rangle\right]
\\ \nonumber 
&&+ \sqrt{\frac{7}{12}}\left[\frac{1}{2}(D^2_{-2,-2}(u)+D^2_{2,-2}(u))\langle 222-2|40\rangle
+D^2_{0,0}(u)\langle 2020|40\rangle\right]
\\ \nonumber
&&+\sqrt{\frac{5}{24}}\left[\frac{1}{2}(D^2_{-2,2}(u)+D^2_{2,2}(u))\langle 2222|44\rangle 
+\frac{1}{2}(D^2_{2,-2}(u)+D^2_{-2,-2}(u))\langle 2-22-2|4-4\rangle\right]
\end{eqnarray}
where the terms with $m_1=\pm 1$ in eq.~\ref{B5} vanish after projection. Upon inserting the Wigner coefficients eq.~\ref{B5b} 
and combining similar terms, we find 
for the lowest polynomial of degree $4$, invariant under the full group $Q\times_s O=S\Gamma$,  the final result
\begin{eqnarray}
\label{B7}
&&\psi^{0,\Gamma_1}=\sqrt{\frac{3}{10}}
\left[\:\frac{1}{2}\left[D^2_{2,2}(u)+D^2_{-2,2}(u)+D^2_{-2,-2}(u)+D^2_{2,-2}(u)\right]
+D^2_{0,0}(u)\right]
\\ \nonumber 
&&= \sqrt{\frac{6}{5}}\left[(x_0^4+x_1^4+x_2^4+x_3^4)
-2(x_0^2x_1^2+x_0^2x_2^2+x_0^2x_3^2+x_1^2x_2^2+x_1^2x_3^2+x_2^2x_3^2)\right].
\end{eqnarray}
The expression in the last line uses eq.~\ref{A1}. It allows to check the invariance both under the deck group $Q$ from Table~\ref{TableN3a} and under the point group $O$.  Eq.~\ref{B7} demonstrates the role of the unimodular Coxeter group $S\Gamma$ as the basis 
underlying {\bf Prop 3}. A similar invariance  under both the deck and the point group
we discuss in section \ref{sec:thetetrahedral} for the tetrahedral 3-manifold.

\end{document}